\documentclass[reprint,preprintnumbers,notitlepage,amsmath,10pt, superscriptaddress,aps,prb]{revtex4-1}
\usepackage{graphicx}
\usepackage{color}
\usepackage{hyperref}
\usepackage{SIunits}
 \usepackage{ulem}
 \usepackage{braket}
\usepackage{subfigure}
\usepackage{mathptmx}
\usepackage{setspace}
\usepackage{anyfontsize}
\usepackage{t1enc}

\def\blue#1{\textcolor{black}{#1}}
\def\red#1{\textcolor{black}{#1}}

\def\Brian#1{\textcolor{black}{#1}}
\def\Tunmi#1{\textcolor{black}{#1}}

\def\bluee#1{\textcolor{black}{#1}}
\def\blue#1{\textcolor{black}{#1}}
\def\bluee#1{\textcolor{black}{#1}}
\def\red#1{\textcolor{black}{#1}}

\def\Brian#1{\textcolor{black}{#1}}
\def\Tunmi#1{\textcolor{black}{#1}}
\def\Peter#1{\textcolor{black}{#1}}

\begin{document}

\title{A strain-tunable quantum dot embedded in a nanowire antenna}

\date{\today}
\author{P. E. Kremer}
 \thanks{P. E. Kremer and A. C. Dada contributed equally to this work.}
\author{A. C. Dada}
 \thanks{P. E. Kremer and A. C. Dada contributed equally to this work.}
\author{P. Kumar}
\author{Y. Ma}
\author{S. Kumar}
\affiliation{Institute of Photonics and Quantum Sciences, SUPA,
Heriot-Watt University, Edinburgh, United Kingdom}

\author{E. Clarke}
\affiliation{EPSRC National Centre for III-V Technologies, University of Sheffield, United Kingdom}

\author{B. D. Gerardot}
 \email{b.d.gerardot@hw.ac.uk}
\affiliation{Institute of Photonics and Quantum Sciences, SUPA,
Heriot-Watt University, Edinburgh, United Kingdom}


\begin{abstract}
We demonstrate an elastically-tunable self-assembled quantum dot in a nanowire antenna that emits single photons with resolution-limited spectral linewidths. The single-photon device is comprised of a single quantum dot embedded in a top-down fabricated nanowire waveguide integrated onto a piezoelectric actuator. Non-resonant excitation leads to static (fluctuating) charges \bluee{likely} at the nanowire surface, causing DC Stark shifts (inhomogeneous broadening); for low excitation powers, the effects are not observed and resolution-limited linewidths are obtained. Despite significant strain-field relaxation in the high-aspect-ratio nanowires, we achieve up to 1.2\hspace{1pt}meV tuning of a dot's transition energy. Single-photon sources with high brightness, resolution-limited linewidths, and wavelength tunability are promising for future quantum technologies.
\end{abstract}

\maketitle
Self-assembled quantum dots (QDs) can generate indistinguishable photons \cite{santori2002indistinguishable, ates2009post, he2013demand}, entangled photon pairs \cite{akopian2006entangled, salter2010entangled}, and entangled spins and photons \cite{deGreve2013complete, gao2012observation, schaibley2013demonstration} due to the large oscillator strengths, clean selection rules, and relatively coherent spin states \cite{press2008complete, brunner2009coherent} of trapped carriers in QDs. To exploit these characteristics for linear-optical quantum computing \cite{knill2001scheme, kok2010introduction} or quantum repeaters and distributed quantum networks \cite{kimble2008quantum,RevModPhys.83.33}, three crucial requirements of scalable QD devices are (i) efficient collection of the spontaneous emission into a single optical mode, (ii) minimal inhomogeneous broadening to enable transform-limited linewidths, and (iii) spectral tunability so that each dot can be made indistinguishable~\cite{patel2010two,PhysRevLett.104.137401}.  

Multiple approaches to enhance the extraction efficiency ($\eta$), defined here as the ratio of power collected by an objective lens to the total power emitted from a dipole, have been pursued.  A common strategy is to create a highly directional far-field radiation pattern, which has been achieved for QDs embedded in both high-Q cavities  \cite{strauf2007high, gazzano2013bright, madsen2014efficient} and low-Q planar cavity structures \cite{davancco2011circular, trotta2012nanomembrane, ma2014efficient}. Recently, sub-wavelength dielectric nanowires have been shown \cite{claudon2010highly, reimer2012bright} to act as highly efficient waveguides with tailorable far-field radiation patterns \cite{gregersen2008controlling, friedler2009solid,claudon2013harnessing}. Unlike high-Q cavities, these waveguides are compatible with large spectral tunability as the spontaneous emission is funnelled into the waveguide over a wide spectral range with high fidelity. Hence, highly tunable and efficient quantum photonic devices can be envisioned with this platform. 

Reversible, \textit{in-situ} manipulation of single particles in QDs can best be achieved with electric \cite{finley2001observation, hogele2004voltage, gerardot2007manipulating, patel2010two,vogel2007influence} and strain \cite{seidl2006effect, trotta2012nanomembrane, kuklewicz2012electro, sapienza2013exciton, PhysRevLett.104.137401} fields. Successful electrical contacting of QDs embedded in vertical nanowires \cite{gregersen2010designs} has yet to be demonstrated due to the difficulty of fabricating reliable nanoscale metal-semiconductor contacts. \textit{In-situ} strain tuning of nanowires also presents challenges not present for bulk structures, as significant strain-field relaxation along the length of the nanowire occurs in high-aspect-ratio structures. Thus far, strain fields have been used to dynamically \bluee{modulate \cite{yeo2013strain,montinaro2014quantum,weiss2014dynamic}} and quasi-permanently control~\cite{bouwes2012controlling} the electronic properties of QDs in nanowire waveguides. However, to date, reversible \textit{in-situ} tuning of two-level emitters in nanowire waveguides has not been realized.

In addition to high $\eta$ and pure single-photon emission, a requirement of single-photon emitters for some applications is transform-limited linewidths ($\Gamma_{rad}$).  However, localized charges in the environment of the QD can shift a dot's emission energy via the quantum confined Stark effect \cite{PhysRevLett.108.107401}. Fluctuations in the microscopic charge distribution in the dot's environment can then lead to spectral fluctuations. The effect of spectral fluctuations is determined by the ratio $\hbar/\Gamma_{rad}$ and the timescale of the charge fluctuation; the spectroscopic manifestation of the fluctuations is also determined by the experimental acquisition time~\cite{berthelot2006unconventional,bounouar2012extraction}. In semiconductors, fluctuating charges are omnipresent, particularly at defects formed at interfaces\cite{PhysRevLett.108.107401} and free-surfaces\cite{wang2004optical,PhysRevB.89.161303}. Therefore, QDs in small-diameter nanowires are particularly susceptible to significant spectral fluctuations~\cite{yeo2011surface,bounouar2012extraction,reimer2014overcoming}. 

\begin{figure*}
  \centering  \includegraphics[width=1.0\linewidth]{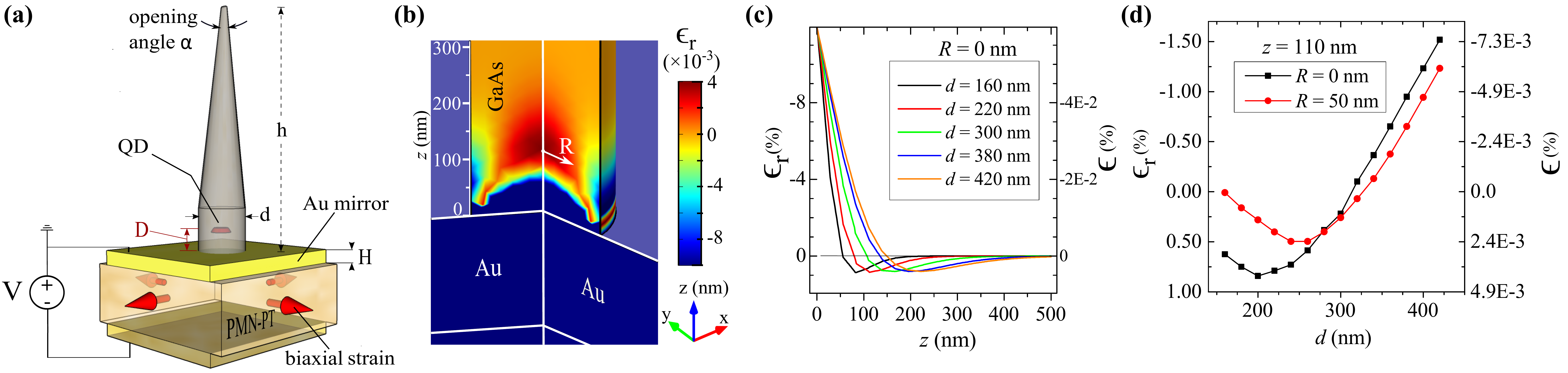}
   \caption{(a) Sketch showing the geometry of our device, where: $D$ is the distance between the quantum dot and the A\Tunmi{u} mirror, $d$ is the diameter of the pillar at the quantum dot position, $h$ is the height of the pillar, and $\alpha$ is the opening angle of the pillar taper. The PMN-PT crystal  has electrical gold contacts (with thickness $H$) on both sides for voltage tuning. The top Au contact also acts as a bonding layer and mirror for the broadband optical antenna. \Tunmi{(b) Simulation of strain relaxation in the nanowire using finite-element method.  The plot shows the profile of relative strain $\epsilon_r = \epsilon(x,y,z)/|\epsilon_0|$, where the strain is $\epsilon(x,y,z)$  and the strain in the PMN-PT crystal is $\epsilon_0 $. The colour legend is scaled to highlight the strain-field relaxation within the nanowire. (c) Plot of the strain $\epsilon$ and relative strain $\epsilon_r$  as a function of the distance along the $z$ axis from the Au/GaAs interface ($z=0\hspace{1pt}$nm at a radially centred position, $R=0\hspace{1pt}$nm).  (d) Plot of $\epsilon$ and  $\epsilon_r$ at $z=110\hspace{1pt}$nm and $R=$0\hspace{1pt}nm,~50\hspace{1pt}nm. The  nanowire diameter $d=220\hspace{1pt}$nm}  in (b), and $\epsilon_0 =- 0.1\%$ in (b)-(d).}
\label{fig:3dsketch}
\end{figure*}

Here we demonstrate an elastically-tunable QD embedded in a nanowire waveguide which emits single photons with linewidths limited by our experimental resolution. We develop a deterministic top-down fabrication procedure to create nanowires with desired geometries. Non-resonant photoluminescence (PL) spectroscopy of QDs in these nanowires shows that the QD emission can be outcoupled with high fidelity, although with less-than-ideal success rates. Statistics from the characterization of 40 QDs in 16 nominally identical nanowires yield $\bar{\eta} = 13 \% \pm 10 \%$, with a maximum $\eta_{max} =57\%$. The large variation in $\eta$ is ascribed to variations in the radial position\Tunmi{s} of the QD\Tunmi{s} as well as surface roughness and asymmetry in the nanowire structures. At low excitation powers, resolution-limited spectral linewidths are found for some QDs. At higher excitation powers, DC Stark shifts and inhomogeneous linewidth broadening are observed due to static and fluctuating electric fields ($F$ and $\delta F$, respectively) at the QD position, respectively. We quantitatively estimate $F$ and $\delta F$ at the QD position by assuming they are generated by filling of nearby nanowire surface states via above-band-gap excitation. The resolution-limited linewidths at low excitation powers lead to optimism that resonant driving of the QDs could be successfully achieved in future experiments for full quantum-optical control of the QD. Such experiments are greatly assisted by \textit{in-situ} tuning of the QD excitonic transition energy~\cite{hogele2004voltage}, which we demonstrate here for the first time with nanowire QDs via bonding to a piezoelectric crystal. Statistics from excitonic transitions in 30 QDs shows \Tunmi{reversible energy-tuning amplitudes ($\delta E$) of $\overline{\delta E} =0.40\pm 0.33$\hspace{1pt}meV (based on tuning-slopes statistics and an applied piezo voltage of 1\hspace{1pt}kV) and a maximum amplitude $\delta E_{max} =1.2$\hspace{1pt}meV}, with small hysteresis effects.  The active strain tuning could enable the reduction of the neutral exciton fine-structure splitting \cite{seidl2006effect, trotta2012nanomembrane, sapienza2013exciton} for the generation of entangled-photon pairs from high-quality self-assembled InGaAs QDs in GaAs nanowires, similar to what has recently been achieved with QDs with nominally small fine-structure splittings in InP nanowires~\cite{huber2014polarization,versteegh2014polarization}.

The design criteria to optimize both the coupling of the QD emission into the fundamental mode of a nanowire waveguide and the directionality of the far-field radiation are well established \cite{friedler2009solid,gregersen2008controlling,claudon2013harnessing}. A {\it reduced nanowire diameter} ($d/ \lambda$) of 0.235 is found to optimally funnel the QD emission into the fundamental $HE_{11}$ mode in both directions along a GaAs nanowire. A mirror terminates one end of the nanowire to reflect incident light towards the outcoupling nanowire end, where a conical taper is introduced to adiabatically expand the confined mode into a plane-wave in free-space. The angle of the conical tapering ($\alpha$) determines both the reflection of the guided mode and the divergence angle of the far-field radiation pattern; for a QD located at the nanowire center, $ \alpha < 10^{\circ}$ leads to $\eta > 50 \%$ 
(\bluee{see} Sec.~SII in the \bluee{Supplemental Material~\footnote{See Supplemental Material at [URL will be inserted by Publisher] for details on the fabrication procedure, determination of the experimental photon extraction efficiency as well as FDTD simulation of the structure.})}. Coupling of the light to the fundamental mode is optimised by placing the dot at the electric field's antinode caused by the standing wave pattern between the mirror and emitter.   
An idealised sketch of our device is shown in Fig.~\ref{fig:3dsketch}~(a). It consists of a self-assembled QD located a distance $D$ from a Au mirror and radially centred in a nanowire with height $h$, diameter $d$, and taper angle $\alpha$. The Au mirror (with thickness $H$ = 200\hspace{1pt}nm) is deposited directly onto a single crystal lead magnesium niobate-lead titanate (PMN-PT) substrate ($ 300\hspace{1pt}\mu$m thick) to also act as an electrical contact for piezoelectric bi-axial strain tuning. \red{This top electrical contact is grounded to prevent large electric fields near the QD.  The complete fabrication procedure is detailed in the \bluee{Supplemental Material~\cite{Note1}}.}  

To better understand the challenge associated with strain tuning a QD in a high-aspect-ratio nanowire, we have simulated the complete device using finite element method (FEM) (Fig.~\ref{fig:3dsketch}~(b)-(d)). 
We quantify strain-field relaxation by means of the {\it relative strain} ($\epsilon_r$), defined as $\epsilon_r= \epsilon(x,y,z)/|\epsilon_0|$, where $\epsilon(x,y,z)$ is the strain at a given position with coordinates $(x,y,z)$, and $\epsilon_0$ is the strain in the PMN-PT crystal. 
  Fig.~\ref{fig:3dsketch}~(b) shows \Tunmi{the profile of relative strain $\epsilon_r$, with $\epsilon_0 = -0.1\%$ and $d=220\hspace{1pt}$nm. The colour legend is scaled to highlight the strain-field relaxation within the nanowire.  In Fig.~\ref{fig:3dsketch}~(c), we show the plot of $\epsilon_r$ (and $\epsilon$) as a function of axial position along the nanowire for different diameters $d$ at the centre of the nanowire,  $R=0\hspace{1pt}$nm. For $D=110\hspace{1pt}$nm, we also show the strain as a function of $d$ for radial positions $R=0\hspace{1pt}$nm,~50\hspace{1pt}nm in Fig.~\ref{fig:3dsketch}~(d).}
 \Tunmi{The modelling first confirmed 
that the strain relaxation is linear with respect to the applied strain, i.e., $\epsilon_r\propto\epsilon_0$, for $\epsilon_0 = -0.05\%, ..., -0.5\%$ (which is within the range expected for a PMN-PT single crystal for an applied voltage of 0\hspace{1pt}-1\hspace{1pt}kV~\cite{herklotz2010electrical,kumar2011strain}). 
 The model shows that the strain field generated by the piezoelectric crystal relaxes substantially ($\approx 80 \%$) across the 200\hspace{1pt}nm-thick Au layer. \bluee{We note that, although the inclusion of a silica spacer between the nanowire and the Au layer has been shown to increase modal reflectivity\cite{gregersen2008controlling,claudon2013harnessing}, its absence in our device enhances strain transfer to the nanowire from the Au layer by $\approx20\%$ based on our simulation results}. The remaining strain field transmitted across the Au/GaAs interface is highly dependent on the diameter of the nanowire, as well as on the radial and axial position within the nanowire. } 
   In particular, we see increased relaxation with reducing nanowire diameter [Fig.~\ref{fig:3dsketch}~(c)] as well as higher strain fields nearer the centre of the nanowire [Fig.~\ref{fig:3dsketch}~(d)].   Although the applied strain is compressive, regions of tensile strain are seen as the strain relaxes along the nanowire.  
The axial position for optimal strain tuning is found to be in conflict with that required for optimal coupling to the $HE_{11} $ mode. In fact, with 
geometry optimised for coupling at $\lambda=950\hspace{1pt}$nm (i.e., $D=80\hspace{1pt}$nm, $d=220\hspace{1pt}$nm, $R=0\hspace{1pt}$nm), we obtain $\epsilon_r=0.04\%$.  While changing to $D<80\hspace{1pt}$nm significantly increases $\epsilon_r$, it will likely lead to increased spectral fluctuations due to the effect of surface states at the mirror/nanowire interface~\cite{PhysRevLett.108.107401, wang2004optical}. 
Keeping $d=220\hspace{1pt}$nm and $R=0\hspace{1pt}$nm for $D>80\hspace{1pt}$nm, $\epsilon_r$ is maximised at $D\approx110\hspace{1pt}$nm before completely relaxing by $D=250\hspace{1pt}$nm.  
For a QD at $ R=0\hspace{1pt}$nm and 
$D=110\hspace{1pt}$nm in a pillar with diameter $d =220\hspace{1pt}$nm, the model predicts a relative strain of $\epsilon_r = 0.8\%$ while $\eta$ is only moderately affected (see \bluee{Supplemental Material~\cite{Note1}}), demonstrating the validity of the device as an efficient and elastically tunable platform for quantum photonics.
\begin{figure*}[ht!]
   \centering  \includegraphics[width=1.0\textwidth]{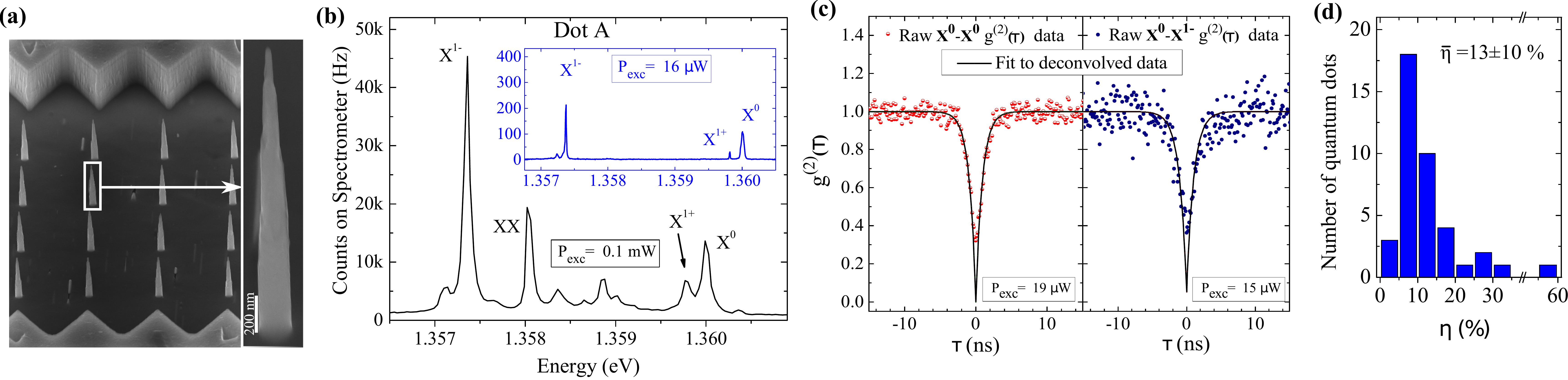}
   \caption{(a) SEM image of an array of nominally identical tapered nanowires with $d = 232\pm4\hspace{1pt}$nm, $D = 110$ nm, $h = 2~\mu$m, and $\alpha = 9.8\pm1.3^{\circ}$. The enlarged image is of the nanowire from which the spectrum in (b) is obtained. \bluee{Unetched areas forming $\mu$-structures ($\approx 3\mu m \times 7\mu m$ in area) are partly visible on the top and bottom edges of the nanowire array}. (b) The PL spectrum from Dot A at saturation power with an estimated \bluee{extraction} efficiency of $\eta=25.9\%$. The inset shows a spectrum from the same dot below saturation. 
  (c) Auto-correlation of $X^{1-}$ and cross-correlation between $X^{1-}$ and $X^{0}$ for Dot B obtained at an excitation power of $19\hspace{1pt}\mu$W. (d) A histogram of the estimated extraction efficiencies $\eta$ from QDs in the 16 nanowires shown in (a).}
\label{fig:Figure2}
\end{figure*}
\vspace{30pt}

We have spectroscopically characterized numerous nanowires with constant $D$ (110\hspace{1pt}nm) and $h$ (2\hspace{1pt}$\mu$m) but varying $d$ and $\alpha$. 
While the nanowires are not deterministically aligned to the randomly positioned QDs, the wafer has a suitably high QD density ($\sim1.2\times10^{10}\hspace{1pt}{\rm cm}^{-2}$) for us to typically observe 3 to 5 spectrally isolated QDs per nanowire.  A scanning-electron-microscope (SEM) image of an array of some of the brightest devices is shown in Fig.~\ref{fig:Figure2} (a), where \blue{$d = 223\hspace{1pt}$nm (corresponding to $0.227 < d/ \lambda < 0.235$ for 920 nm $< \lambda <$ 950 nm,} the typical range of ground state exciton emission wavelengths for the QDs in this wafer),  and $\alpha \approx 10^{\circ}$.  

PL spectra were acquired from the sample at $T = 4.5\hspace{1pt}$K using a confocal microscope with a 0.82 NA objective lens. The QDs were excited non-resonantly ($\lambda_{exc}=830\hspace{1pt}$nm) using a continuous-wave (CW) laser diode. To maximize the PL signal we focused the collection optics at the top of the nanowire while the excitation laser was focused at the bottom of the nanowire. The PL was characterized using a spectrometer (0.5\hspace{1pt}m focal length and $1800\hspace{2pt}{\rm grooves\hspace{2pt}mm}^{-1}$ grating giving a resolution of $\Gamma_{\rm spect}=35\hspace{1pt}\mu$eV [full width at half maximum (FWHM)] measured using a narrow-band laser) and a 
liquid-nitrogen-cooled Si charge-coupled-device (CCD) camera. Second-order correlation measurements were acquired using a Hanburry Brown-Twiss setup ($350\hspace{1pt}\mu$eV spectral resolution) with two silicon single-photon avalanche diodes (timing jitter $\approx 500$ ps) and timing electronics.

The PL spectra from a QD (Dot A) found in one of the brightest wires is shown in Fig.~\ref{fig:Figure2} (b). The exciton states are identified by the characteristic Coulomb interactions observed in experiments with similar QDs in charge-tunable devices~\cite{PhysRevB.77.245311} and the linear (quadratic) power dependence for single (bi-) excitons. The spectra in Fig.~\ref{fig:Figure2} (b) \bluee{are} representative of what we typically observe from QDs in the nanowires, except with varying central energies and peak intensities. Further confirmation of the state assignment can be obtained by measuring the photon-intensity correlations between separate excitonic states~\cite{PhysRevLett.87.257401}. Figure~\ref{fig:Figure2} (c) shows the second order auto-correlation measurement of the $X^{1-}$ state from another QD, Dot B. The raw data shows $g^{2}(0) <$ 0.5 while the fit to data deconvolved for detector jitter shows $g^{2}(0) \approx 0$, signifying high-purity single-photon emission. The second order cross-correlation measurement between the $X^{0}$ and $X^{1-}$ states (Fig.~\ref{fig:Figure2} (d)) also demonstrates clear antibunching, signifying that these states originate from the same QD \cite{gerardot2005coupledqds}. \bluee{Also, the absence of complex dynamics in the second-order correlation experiments demonstrates charge-state stability on short timescales~\cite{PhysRevB.89.161303,PhysRevB.69.205324}.}

The correct excitonic line assignment \bluee{enables} estimation of the total $\eta$ from a single dot, defined as the power collected into the objective lens divided by the power emitted by the QD. The latter can be estimated at saturation from the emission rates of each excitonic state (defined as the inverse of the transition's lifetime) and the relative integrated intensities of each excitonic state. The efficiency of the experimental setup (from the objective lens to the CCD camera) was calibrated using a tunable laser at the QD emission wavelength. $\eta$ was thus estimated for the 40 brightest dots in the 16 nominally identical nanowires, as summarised in the histogram in Fig.~\ref{fig:Figure2}~(d). We find $\eta_{max} = 57 \%$ for our brightest dot when each excitonic state is included, and $\bar{\eta} = 13 \% \pm 10 \%$. 
The most obvious explanation for the large standard deviation of $\eta$ is the random positioning of the dots radially in the nanowires, which could be remedied in the future with a deterministic positioning technique \cite{gazzano2013bright}. Additionally, the SEM images reveal surface defects and slight asymmetry in the nanowires, which may adversely affect performance. However, we are unable to correlate differences in particular structures with their brightness. Curiously, we have spectroscopically characterized numerous nanowire structures made from the same wafer with smoother features and better symmetry but worse $\eta$.   

Non-resonant PL spectroscopy enables investigation of the effect of the nearby nanowire surfaces on the spectral linewidths and exciton energies of QDs. The non-resonant laser excites carriers above the GaAs band-gap that can relax into the QD as well as fill defect surface states; increasing the excitation power ($P_{exc}$) increases the number of carriers at the nanowire surface. For the self-assembled InGaAs QDs investigated here, $\Gamma_{rad} \sim1\hspace{1pt}\mu$eV. Figs 3 (a) and (b) show the energy detuning ($\Delta E$) and measured linewidths ($\Gamma$) for the $X^{1-}$ for Dots A and B, respectively. For $P_{exc} \leq 20\hspace{1pt}\mu$W for Dot A, $\Delta E$ is zero and $\Gamma$ is resolution limited. For Dot B, $\Delta E$ is zero and $\Gamma$ is constant ( $ \sim45\hspace{1pt}\mu$eV) but inhomogeneously broadened above the experimental resolution limit in the low power regime ($P_{exc} \leq 100\hspace{1pt}\mu$W).  
For both QDs, as $P_{exc}$ increases from the low power regime, $\Delta E$ increases and $\Gamma$ broadens without saturation. To better understand and quantify the effect of the surface states, we exploit the ability of the QD itself to function as an {\it in-situ} probe of the local electric field~\cite{gerardot2007manipulating,vogel2007influence,PhysRevLett.108.107401,PhysRevLett.107.166802}. In an electric field, the quantum dot dipole manifests a Stark shift with quadratic field dependence, and the energy detuning is given by $\Delta E = -pF+\beta F^2$, where $F$ is the electric field, $p$ the permanent dipole moment, and $\beta$ \bluee{the} polarizability. 
 \begin{figure}[h!]
   \centering
\includegraphics[width=1.0\linewidth]{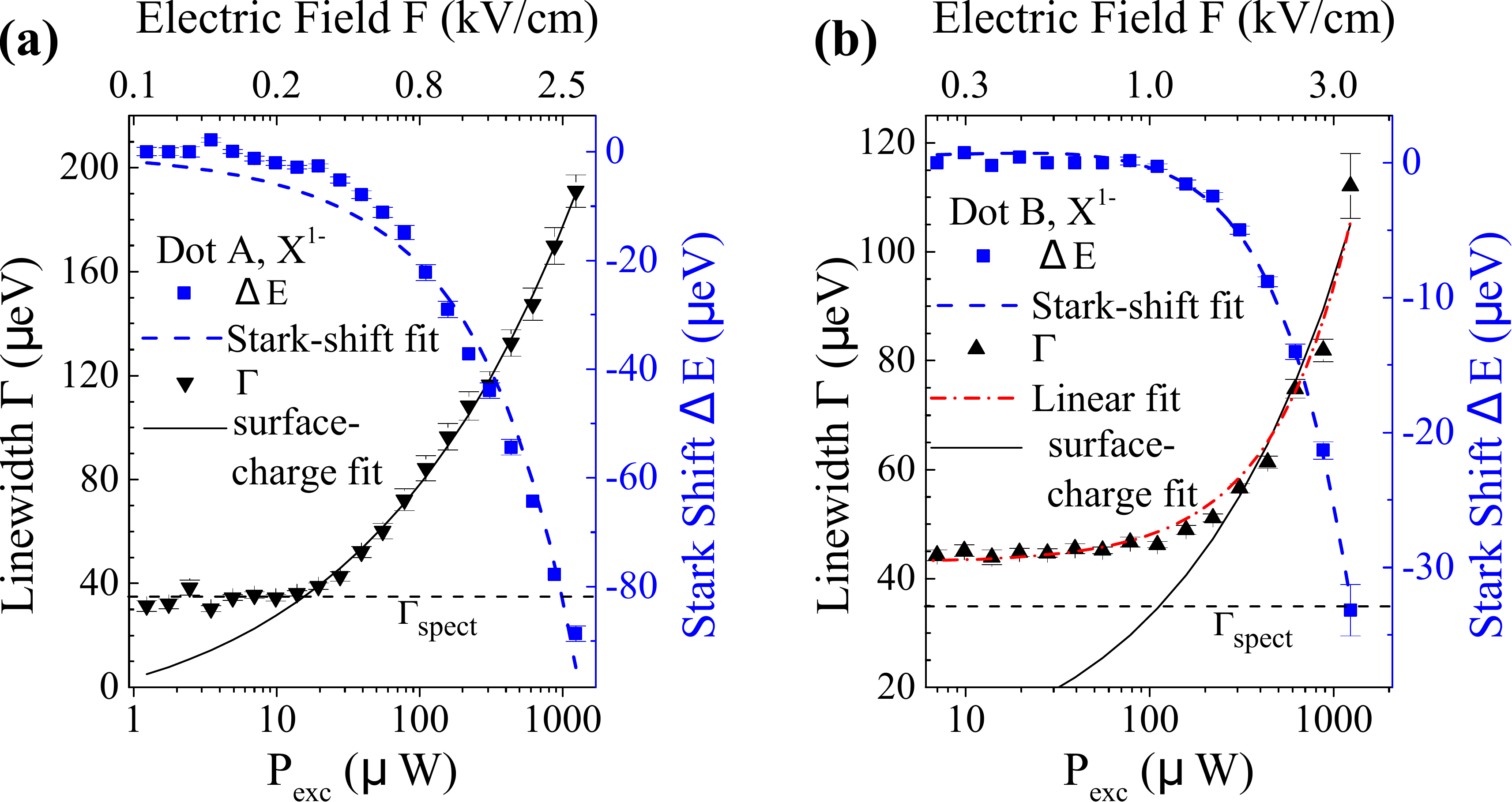}
   \caption{The measured spectral linewidths ($\Gamma$) and Stark shifts ($\Delta E$) as a function of excitation power ($P_{exc}$) are shown for the $X^{1-}$ line for Dot A and Dot B in (a) and (b), respectively. A fit of the Stark effect allows an estimated electric field ($F$) to be calculated in each case. For Dot A, the \textit{surface-charge fit} used to describe the dependence of $\Gamma$ on $P_{exc}$ is based on a fluctuating $\delta F$ estimated from the electronic shot noise of the static $F$. For Dot B, the \textit{surface-charge fit} fails to match the data as $\Gamma$ exhibits a linear dependence to $P_{exc}$.}
\label{fig:fig3}
\end{figure}
\bluee{To gain physical insight, we express $p=er$, where $e$ is the electronic charge, and $r$ is the electron-hole wave function separation.}

 To estimate $F$ at the position of the QD based on $\Delta E$ observed in PL measurements, we assume $\beta=-4\hspace{1pt}\mu{\rm eV/(kV/cm)^2}$~\cite{gerardot2007manipulating,vogel2007influence}. The fits from this procedure \bluee{give $r=2.3\pm0.6$\AA~and $-0.34\pm0.02$\AA~for dots A and B respectively. The fits} are shown as the dashed lines in Fig.~\ref{fig:fig3}, with the extracted values for $F$ shown on the top $x$-axis in the figures. The fits agree well with the data and \bluee{we observe that} $F$ is linearly proportional to $\sqrt{P_{exc}}$ for each QD. Further, we can estimate the fluctuating electric field, $\delta F$, by assuming the  fluctuation ($\delta n$) in the number of electrons ($n$) located at the surface is proportional to the electronic shot noise: $\delta n \propto \sqrt{n}$. We calculate $n$  by assuming electrons a distance $d$/2 create $F$, and then use $\delta F$ to find the corresponding $\delta(\Delta E)\equiv\Gamma$ from the Stark equation. 
 For Dot A, the estimated power broadening of $\Gamma$ (shown as the \Tunmi{solid line} labeled \textit{surface-charge fit} in Fig. 3(a)) fits the experimental data very well above the system's resolution limit. The surface-charge fit also enables us to estimate the power broadening below the resolution limit. Both the data and the extrapolated fit suggest the inhomogeneous broadening is dominated by the charges generated by the excitation laser at the nanowire surface. The results from Dot A are promising for the generation of indistinguishable photons from a QD in a nanowire antenna. On the other hand, a fit with a linear dependence of $\Gamma$ on $P_{exc}$ is found to fit the data for Dot B much better than a surface-charge fit, even in the high $P_{exc}$ regime. This result suggests that unlike the behavior of Dot A's power broadening, Dot B's linewidth broadening \bluee{is not solely caused by the charge fluctuations generated at the nanowire surface by non-resonant excitation}.  

\begin{figure}[t]
   \centering
 \includegraphics[ width=1.0\linewidth]{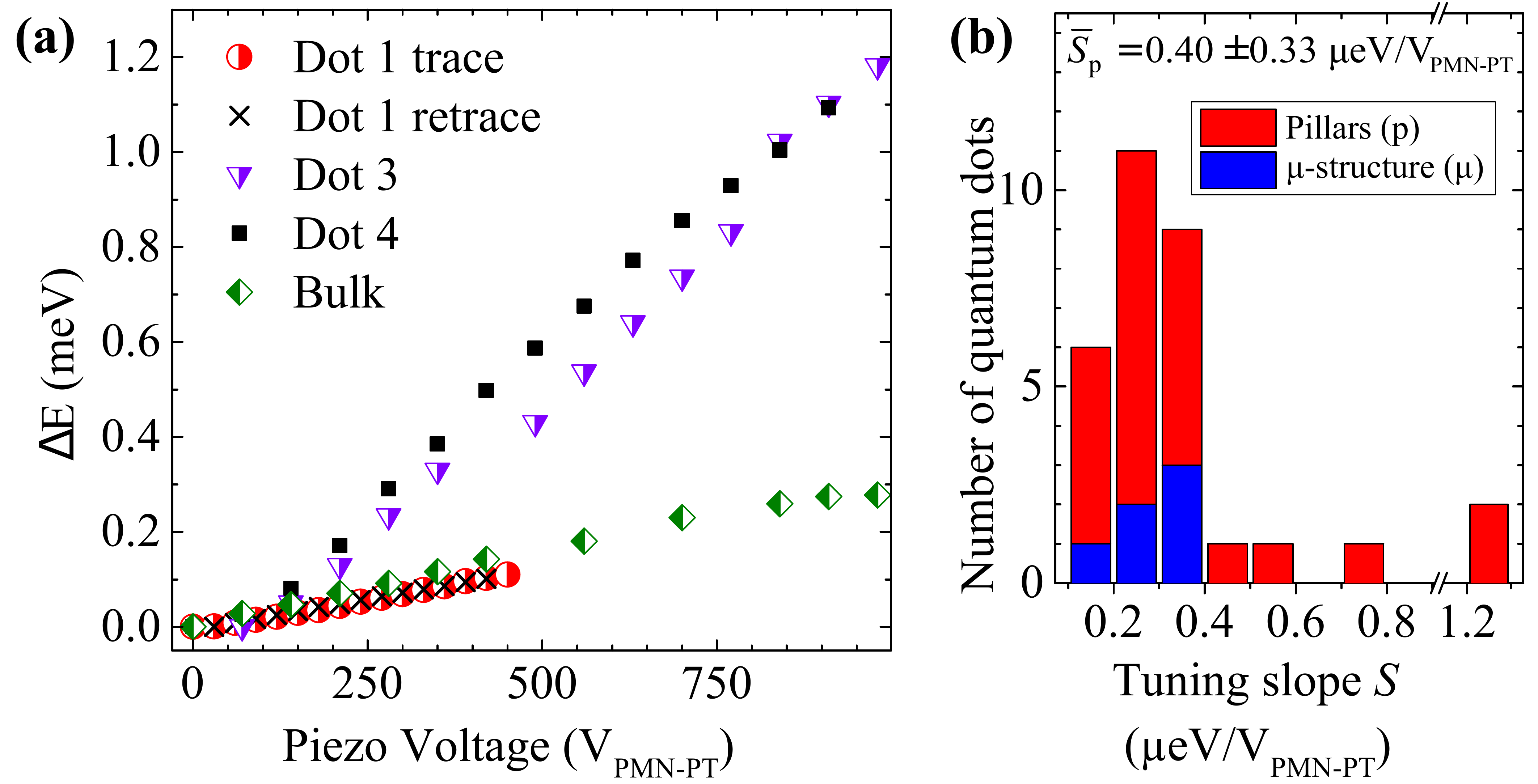}
   \caption{(a) Strain tuning the energies of different single QD excitons. Each QD in each pillar exhibits a different strain tuning slope ($S=\Delta E / \Delta V_{\rm PMN-PT}$), as shown in the histogram in (b). Also shown in the histogram are tuning slopes $S_{\rm b}$ for QDs in the \bluee{$\mu$-structure membrane ($S_{\rm \mu}=0.29\pm0.06\hspace{1pt}\mu \rm{eV/  V}_{\rm PMN-PT}$)}.}
\label{fig4:Figure4}
\end{figure} 

Finally, we demonstrate the elastic tunability of the exciton states. Fig.~\ref{fig4:Figure4} (a) shows several examples of exciton detuning for QDs in different nanowire antennas as a function of applied bias to the piezoelectric crystal ($V_{PMN-PT}$). We observe varied strain-tuning slopes for each QD in each nanowire, as summarised for 30 QDs in the histogram in Fig.~\ref{fig4:Figure4} (b). Also shown is a histogram of the strain tuning of QDs located in the \bluee{roughly $3\mu m \times 7\mu m$} unetched region of the sample nearby the nanowires \bluee{($\mu$-structure). This is partly visible in Fig.~\ref{fig:Figure2} (a) on the top and bottom edges of the array}. In general, the strain tuning is achieved with small amounts of hysteresis, as shown for \Tunmi{Dot 1}. A maximum tuning amplitude of $\delta E\approx1.2\hspace{1pt}$meV was achieved in the experiment. The large standard deviation in tuning is expected for two reasons: (i) the amplitude and even sign of the strain field is highly dependent on the radial position of the QD in the nanowire (as shown in Fig.~\ref{fig:3dsketch}); and (ii) strain tuning of the quantum states is highly dependent on the exact morphology of the dot, which is unique for every QD~\cite{sapienza2013exciton,kuklewicz2012electro,jons2011dependence}. In spite of the significant strain-field relaxation in the Au contact and the nanowire, we 
achieve substantial {\it in-situ} strain tuning of the QD excitonic transition energies. Further improvement in the tuning range can be obtained by reducing the Au layer thickness $H$ and moving the QD closer to the mirror, which may however lead to more spectral fluctuations.

In conclusion,  we have demonstrated an elastically-tunable QD embedded in a nanowire waveguide emitting single photons with resolution-limited linewidths. 
The device enables strain tuning of excitons by up to 1.2\hspace{1pt}meV, which could enable resonant fluorescence experiments, reduction of fine-structure splitting for entangled photon-pair generation, and two-photon interference from separate QDs in nanowire antennas.

{\bf Acknowledgements:} The authors would like to thank G.C. Ballesteros for assistance with FEM software (COMSOL), and acknowledge the financial support for this work from a Royal Society University Research Fellowship, the EPSRC (\bluee{grant numbers: EP/I023186/1, EP/K015338/1}), and an ERC Starting Grant (number: 307392).  


%


\newpage

\renewcommand{\thefigure}{S\arabic{figure}}
\renewcommand{\theequation}{S\arabic{equation}}
\renewcommand{\thesection}{S\Roman{section}}
\renewcommand{\thesubsection}{s\Roman{section}}
\renewcommand{\thetable}{S\arabic{table}}

\setcounter{figure}{0}
\setcounter{equation}{0}
\setcounter{section}{0}

\onecolumngrid 
\centering
\newpage
{\bf \Large Supplemental Material for \\``A strain-tunable quantum dot embedded in a nanowire antenna''}
\flushleft
\onecolumngrid
\section{Fabrication procedure}

{\fontsize{12}{20}\selectfont
Our fabrication procedure is illustrated in Fig.~\ref{fig1:aFabschem}. We use a wafer grown by molecular beam epitaxy (MBE) consisting of a $1000$\hspace{1pt}nm Al$_{0.65}$Ga$_{0.35}$As sacrificial-etch layer followed by a $2\hspace{1pt}\mu$m thick GaAs layer in which a layer of QDs is grown 110\hspace{1pt}nm from the surface. A 100\hspace{1pt}nm (Au) mirror is deposited by electron beam evaporation. After this, the wafer is flipped and bonded to a PMN-PT crystal with a 100\hspace{1pt}nm-thick Au layer using a thermo-compression bonding step at temperature and pressure $T = 300^{\circ}\hspace{1pt}  {\rm C}$ and $P = 2$\hspace{1pt}MPa respectively. Hydrochloric acid at $T = 0^{\circ}\hspace{1pt}{\rm C}$ is used to selectively etch the sacrificial layer, allowing removal of the substrate wafer. Next, an e-beam-lithography shadow mask is defined in a 220\hspace{1pt}nm layer of PMMA/MMA [Poly(methyl methacrylate)] for the 90\hspace{1pt}nm-thick Ni dry-etch mask deposited by e-beam evaporation. 
\begin{figure*}[h!b!]
   \centering
  \includegraphics[width=\linewidth]{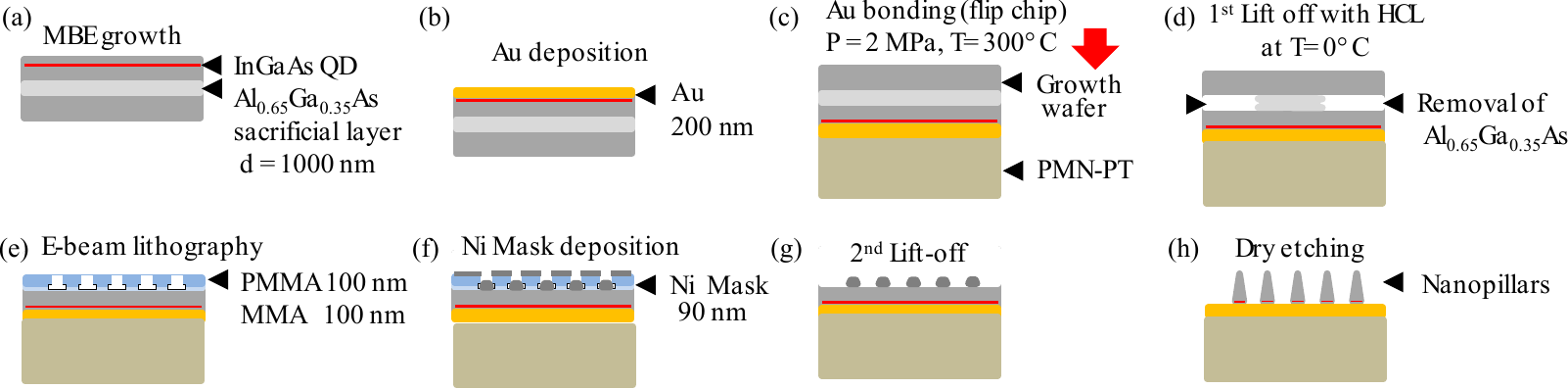}
   \caption{ A schematic of the fabrication procedure. (a) A sample consisting of self-assembled InGaAs quantum dots with a 110\hspace{1pt}nm-thick capping layer and embedded in a 2 $\mu$m GaAs layer on an Al$_{0.65}$Ga$_{0.35}$As sacrificial-etch layer is grown by MBE. (b) A 100 nm Au back mirror is deposited. (c) Following a flip-chip process, the Au layer is attached to the Au-coated PMN-PT crystal using thermo-compression bonding ($T = 300$\hspace{1pt}C$^{\circ}$ and $P = 2$ MPa). (d) Hydrochloric acid at $T = 0^{\circ} C$ is used to selectively etch the Al$_{0.65}$Ga$_{0.35}$As layer. (e) Electron beam lithography is used to define circular apertures of the desired radius. (f) 90 nm of Ni is deposited followed by (g) lift-off and (f) dry  etching of nanowires. }
   \label{fig1:aFabschem}
\end{figure*}
The  PMMA/MMA is removed using acetone in an ultrasonic bath to expose the Ni discs which act as masks for the final dry-etching process.
 
The dry-etching was carried out on an Oxford Plasmalab 100 ICP 65 system using a two-step process. A recipe for a straight etch (Tab.~\ref{table:Table1} col.~(j)) followed by a recipe for controlled undercut of the mask (\bluee{Table}~\ref{table:Table1} col. (k))  resulted in the nanowires characterized in the manuscript. 

\begin{table}[h] 
\begin{tabular}{|l|c|c|}
\hline
~~~Parameter @ 20\hspace{1pt}$^{\circ}C$&~~~~~~Straight (j)~~~~~~&~~~~~~Cone (k)~~~~~~\\
\hline
RF & 60 W & 90 W \\
\hline
ICP&200 W&150 W\\
\hline
Pressure & 8 mT & 2.7 mT\\
\hline
SiCl$_{4}$/Ar & 7.5/15 sccm & 4/12 sccm\\
\hline
\end{tabular}
\caption {\large Etch parameters for straight (j) and cone shaped (k) wires.}
\label{table:Table1}
\end{table}To remove the remaining Ni and clean the taper from redeposited material, the sample was immersed in  1:10 Nitric acid:H$_{2}$O for 2 minutes. The nitric acid treatment improved the taper angle by up to $ 2^{\circ}$. Fig.~\ref{fig1:fabsem} shows SEM images demonstrating the influence of dry etch parameters on the nanowire shape, i.e., RF power and RIE (reactive ion etch) gas chamber pressure.

\begin{figure*}[b]
   \centering
  \includegraphics[width=\linewidth]{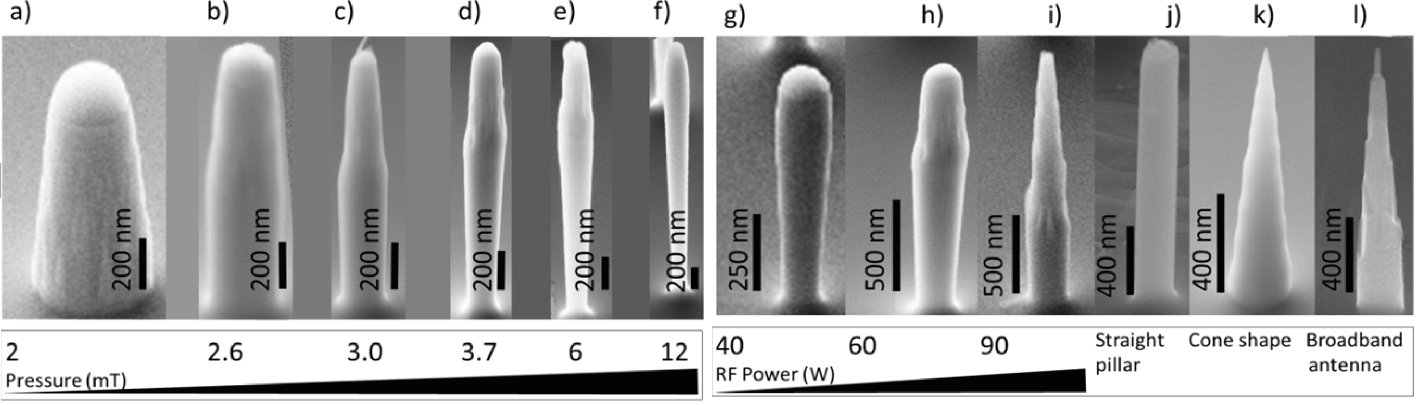}
   \caption{ SEM images showing the influence of dry etch parameters on the nanowire shape, i.e., RF power, and RIE (reactive ion etch) gas chamber pressure.}
   \label{fig1:fabsem}
\end{figure*}

\bluee{\section{Photon extraction  efficiency}}

We perform finite-difference time-domain (FDTD) simulations (using Lumerical\textsuperscript{\textregistered}) to model the performance of the nanowire antenna. The nanowire is characterised in terms of the following figures of merit: (1) The Purcell factor $P_m$, which describes the enhancement or suppression of the spontaneous emission rate of the QD into a guided mode of the waveguide as compared with emission in bulk material. The Purcell factor can be moderately enhanced by adjusting the waveguide diameter to QD emission wavelength ratio $d/\lambda$ and the dipole to mirror distance $D$ ~\cite{friedler2009solid}. (2) The mode-coupling efficiency or $\beta$ factor, which describes the fraction of photons emitted into the outcoupled mode in comparison with the total emission into all modes. 
\begin{figure}[h!]
   \centering  
 \includegraphics[ width=0.55\linewidth]{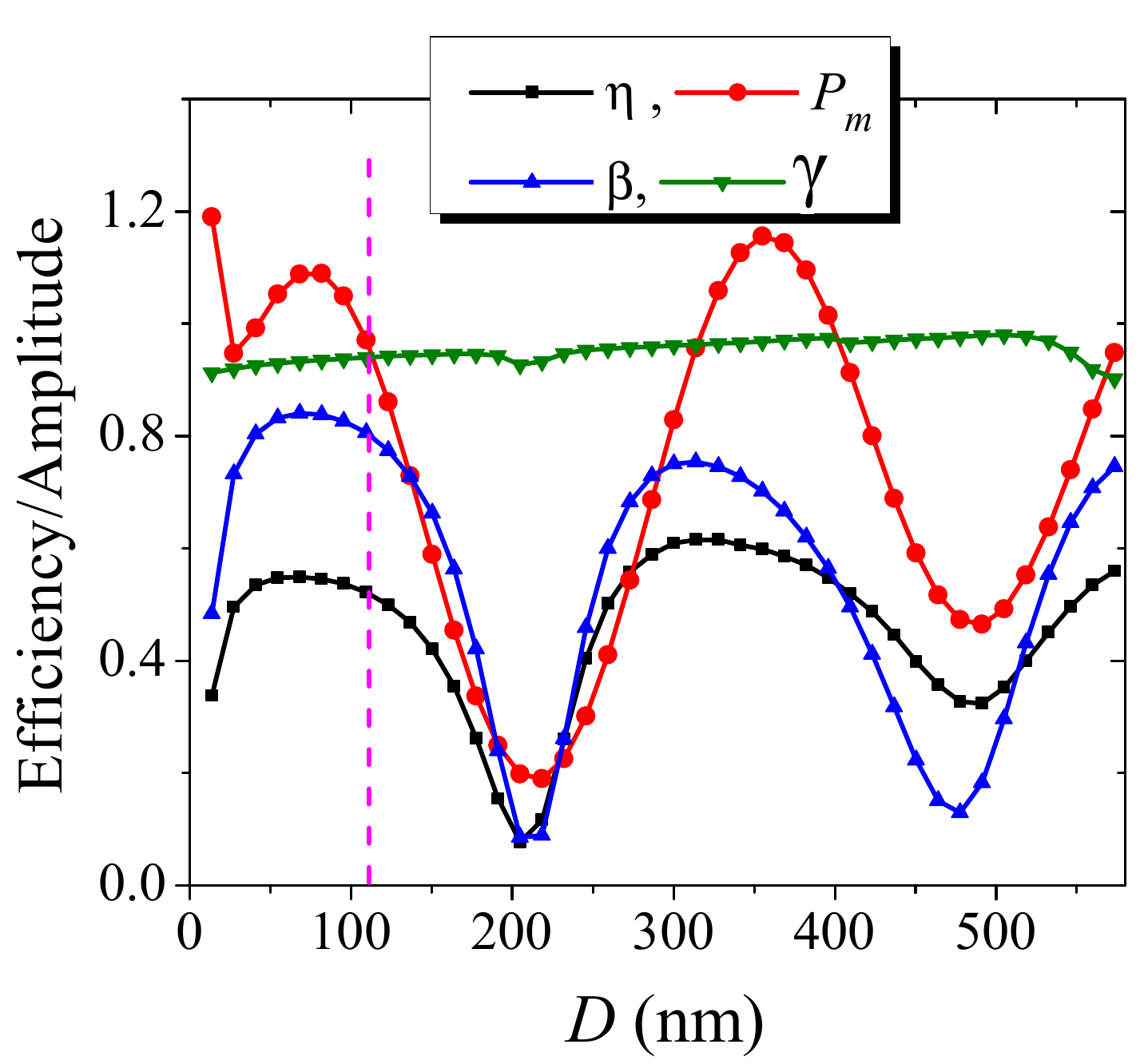} 
    \caption{FDTD simulation results showing $\eta$, ${P}_m$, $\beta$  and  $\Gamma$  of a dipole a distance $D$ from the Au back mirror. For this simulation, $\alpha=10{^\circ}$, $\lambda$ = 950 nm, and $d/\lambda=0.235$ and the dipole is centered radially. The dashed line corresponds to $D=110\hspace{1pt}$nm, the value used in our devices.}
    \label{fig:FDTD2}
\vspace{10pt}
   \centering  
   \includegraphics[ width=0.65\linewidth]{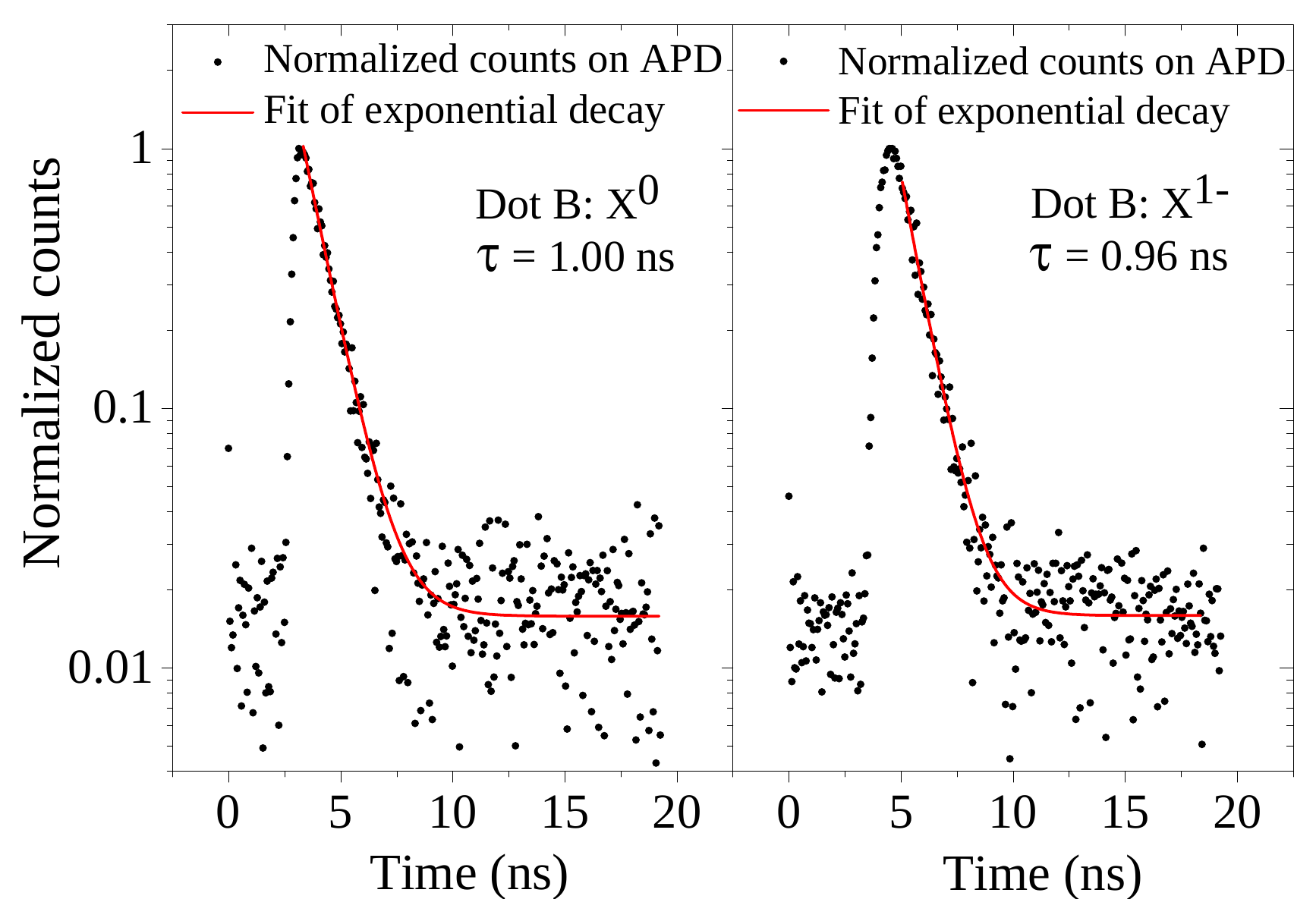}
  \caption{Time-resolved photoluminescence of the $X{^0}$ and $X^{1-}$ lines from Dot B measured using an avalanche photodiode (APD). }
     \label{fig:Life3}
\end{figure} 
In the case of $d/\lambda=0.235$ the only supported mode is the fundamental $HE_{11}$ mode. (3) The extraction efficiency $\eta$, which we define as the ratio of power emitted by the dipole and the power collected by the objective lens (with NA = 0.82 in our setup). (4) \Peter{ The collection efficiency\hspace{1pt} $\gamma$,} which is the ratio of light coupled out of the nanowire to that collected by the objective lens.  Fig.~\ref{fig:FDTD2} shows $\eta$, $\beta$, $P_m$, and $\gamma$ as functions of $D$ for $d=220\hspace{1pt}$nm, $h=2\hspace{1pt}\mu$m, and $\alpha=10^\circ$, and an objective lens with NA of 0.82.

To experimentally determine the extraction efficiency $\eta$ of the nanowire (defined as the power collected into the objective lens divided by the power emitted by the QD), time-correlated single photon counting measurements of the excitonic transitions were performed. As mentioned in the main text, the power emitted by the QD at saturation is estimated from the emission rates of each excitonic state (defined as the inverse of the transition's lifetime) and their relative integrated intensities. \bluee{We assume ideal quantum efficiency for the QD. Our time-resolved measurements (see Fig.~\ref{fig:Life3} and Fig.~2\hspace{1pt}(b) of the main manuscript) show charge state stability over short time scales and no evidence of complex dynamics, such as bi-exponential decay or bunching in autocorrelation measurements~\cite{PhysRevB.69.205324,PhysRevB.89.161303}, while photoluminescence spectra demonstrate charge state stability over long time scales}. 
 The measurements of the $X{^0}$ and $X^{1-}$ transitions for \Brian{Dot B} shown in Fig.~\ref{fig:Life3} each reveal transition lifetimes of \bluee{$\tau \approx 1$\hspace{1pt}ns}, approximately 10\% to 20\% longer than statistically expected for such dots \cite{PhysRevB.77.245311}. \bluee{This is} in agreement with the slight inhibition of spontaneous emission predicted by $P_{m}$ in Fig.~\ref{fig:FDTD2}.


\begin{thebibliography}{57}%
\makeatletter
\providecommand \@ifxundefined [1]{%
 \@ifx{#1\undefined}
}%
\providecommand \@ifnum [1]{%
 \ifnum #1\expandafter \@firstoftwo
 \else \expandafter \@secondoftwo
 \fi
}%
\providecommand \@ifx [1]{%
 \ifx #1\expandafter \@firstoftwo
 \else \expandafter \@secondoftwo
 \fi
}%
\providecommand \natexlab [1]{#1}%
\providecommand \enquote  [1]{``#1''}%
\providecommand \bibnamefont  [1]{#1}%
\providecommand \bibfnamefont [1]{#1}%
\providecommand \citenamefont [1]{#1}%
\providecommand \href@noop [0]{\@secondoftwo}%
\providecommand \href [0]{\begingroup \@sanitize@url \@href}%
\providecommand \@href[1]{\@@startlink{#1}\@@href}%
\providecommand \@@href[1]{\endgroup#1\@@endlink}%
\providecommand \@sanitize@url [0]{\catcode `\\12\catcode `\$12\catcode
  `\&12\catcode `\#12\catcode `\^12\catcode `\_12\catcode `\%12\relax}%
\providecommand \@@startlink[1]{}%
\providecommand \@@endlink[0]{}%
\providecommand \url  [0]{\begingroup\@sanitize@url \@url }%
\providecommand \@url [1]{\endgroup\@href {#1}{\urlprefix }}%
\providecommand \urlprefix  [0]{URL }%
\providecommand \Eprint [0]{\href }%
\providecommand \doibase [0]{http://dx.doi.org/}%
\providecommand \selectlanguage [0]{\@gobble}%
\providecommand \bibinfo  [0]{\@secondoftwo}%
\providecommand \bibfield  [0]{\@secondoftwo}%
\providecommand \translation [1]{[#1]}%
\providecommand \BibitemOpen [0]{}%
\providecommand \bibitemStop [0]{}%
\providecommand \bibitemNoStop [0]{.\EOS\space}%
\providecommand \EOS [0]{\spacefactor3000\relax}%
\providecommand \BibitemShut  [1]{\csname bibitem#1\endcsname}%
\let\auto@bib@innerbib\@empty
\bibitem [{\citenamefont {Santori}\ \emph {et~al.}(2002)\citenamefont
  {Santori}, \citenamefont {Fattal}, \citenamefont {Vu{\v{c}}kovi{\'c}},
  \citenamefont {Solomon},\ and\ \citenamefont
  {Yamamoto}}]{santori2002indistinguishable}%
  \BibitemOpen
  \bibfield  {author} {\bibinfo {author} {\bibfnamefont {C.}~\bibnamefont
  {Santori}}, \bibinfo {author} {\bibfnamefont {D.}~\bibnamefont {Fattal}},
  \bibinfo {author} {\bibfnamefont {J.}~\bibnamefont {Vu{\v{c}}kovi{\'c}}},
  \bibinfo {author} {\bibfnamefont {G.~S.}\ \bibnamefont {Solomon}}, \ and\
  \bibinfo {author} {\bibfnamefont {Y.}~\bibnamefont {Yamamoto}},\ }\href@noop
  {} {\bibfield  {journal} {\bibinfo  {journal} {Nature}\ }\textbf {\bibinfo
  {volume} {419}},\ \bibinfo {pages} {594} (\bibinfo {year}
  {2002})}\BibitemShut {NoStop}%
\bibitem [{\citenamefont {Ates}\ \emph {et~al.}(2009)\citenamefont {Ates},
  \citenamefont {Ulrich}, \citenamefont {Reitzenstein}, \citenamefont
  {L{\"o}ffler}, \citenamefont {Forchel},\ and\ \citenamefont
  {Michler}}]{ates2009post}%
  \BibitemOpen
  \bibfield  {author} {\bibinfo {author} {\bibfnamefont {S.}~\bibnamefont
  {Ates}}, \bibinfo {author} {\bibfnamefont {S.}~\bibnamefont {Ulrich}},
  \bibinfo {author} {\bibfnamefont {S.}~\bibnamefont {Reitzenstein}}, \bibinfo
  {author} {\bibfnamefont {A.}~\bibnamefont {L{\"o}ffler}}, \bibinfo {author}
  {\bibfnamefont {A.}~\bibnamefont {Forchel}}, \ and\ \bibinfo {author}
  {\bibfnamefont {P.}~\bibnamefont {Michler}},\ }\href@noop {} {\bibfield
  {journal} {\bibinfo  {journal} {Physical Review Letters}\ }\textbf {\bibinfo
  {volume} {103}},\ \bibinfo {pages} {167402} (\bibinfo {year}
  {2009})}\BibitemShut {NoStop}%
\bibitem [{\citenamefont {He}\ \emph {et~al.}(2013)\citenamefont {He},
  \citenamefont {He}, \citenamefont {Wei}, \citenamefont {Wu}, \citenamefont
  {Atat{\"u}re}, \citenamefont {Schneider}, \citenamefont {H{\"o}fling},
  \citenamefont {Kamp}, \citenamefont {Lu},\ and\ \citenamefont
  {Pan}}]{he2013demand}%
  \BibitemOpen
  \bibfield  {author} {\bibinfo {author} {\bibfnamefont {Y.-M.}\ \bibnamefont
  {He}}, \bibinfo {author} {\bibfnamefont {Y.}~\bibnamefont {He}}, \bibinfo
  {author} {\bibfnamefont {Y.-J.}\ \bibnamefont {Wei}}, \bibinfo {author}
  {\bibfnamefont {D.}~\bibnamefont {Wu}}, \bibinfo {author} {\bibfnamefont
  {M.}~\bibnamefont {Atat{\"u}re}}, \bibinfo {author} {\bibfnamefont
  {C.}~\bibnamefont {Schneider}}, \bibinfo {author} {\bibfnamefont
  {S.}~\bibnamefont {H{\"o}fling}}, \bibinfo {author} {\bibfnamefont
  {M.}~\bibnamefont {Kamp}}, \bibinfo {author} {\bibfnamefont {C.-Y.}\
  \bibnamefont {Lu}}, \ and\ \bibinfo {author} {\bibfnamefont {J.-W.}\
  \bibnamefont {Pan}},\ }\href@noop {} {\bibfield  {journal} {\bibinfo
  {journal} {Nature Nanotechnology}\ }\textbf {\bibinfo {volume} {8}},\
  \bibinfo {pages} {213} (\bibinfo {year} {2013})}\BibitemShut {NoStop}%
\bibitem [{\citenamefont {Akopian}\ \emph {et~al.}(2006)\citenamefont
  {Akopian}, \citenamefont {Lindner}, \citenamefont {Poem}, \citenamefont
  {Berlatzky}, \citenamefont {Avron}, \citenamefont {Gershoni}, \citenamefont
  {Gerardot},\ and\ \citenamefont {Petroff}}]{akopian2006entangled}%
  \BibitemOpen
  \bibfield  {author} {\bibinfo {author} {\bibfnamefont {N.}~\bibnamefont
  {Akopian}}, \bibinfo {author} {\bibfnamefont {N.~H.}\ \bibnamefont
  {Lindner}}, \bibinfo {author} {\bibfnamefont {E.}~\bibnamefont {Poem}},
  \bibinfo {author} {\bibfnamefont {Y.}~\bibnamefont {Berlatzky}}, \bibinfo
  {author} {\bibfnamefont {J.}~\bibnamefont {Avron}}, \bibinfo {author}
  {\bibfnamefont {D.}~\bibnamefont {Gershoni}}, \bibinfo {author}
  {\bibfnamefont {B.~D.}\ \bibnamefont {Gerardot}}, \ and\ \bibinfo {author}
  {\bibfnamefont {P.~M.}\ \bibnamefont {Petroff}},\ }\href@noop {} {\bibfield
  {journal} {\bibinfo  {journal} {Physical Review Letters}\ }\textbf {\bibinfo
  {volume} {96}},\ \bibinfo {pages} {tunable} (\bibinfo {year}
  {2006})}\BibitemShut {NoStop}%
\bibitem [{\citenamefont {Salter}\ \emph {et~al.}(2010)\citenamefont {Salter},
  \citenamefont {Stevenson}, \citenamefont {Farrer}, \citenamefont {Nicoll},
  \citenamefont {Ritchie},\ and\ \citenamefont
  {Shields}}]{salter2010entangled}%
  \BibitemOpen
  \bibfield  {author} {\bibinfo {author} {\bibfnamefont {C.}~\bibnamefont
  {Salter}}, \bibinfo {author} {\bibfnamefont {R.}~\bibnamefont {Stevenson}},
  \bibinfo {author} {\bibfnamefont {I.}~\bibnamefont {Farrer}}, \bibinfo
  {author} {\bibfnamefont {C.}~\bibnamefont {Nicoll}}, \bibinfo {author}
  {\bibfnamefont {D.}~\bibnamefont {Ritchie}}, \ and\ \bibinfo {author}
  {\bibfnamefont {A.}~\bibnamefont {Shields}},\ }\href@noop {} {\bibfield
  {journal} {\bibinfo  {journal} {Nature}\ }\textbf {\bibinfo {volume} {465}},\
  \bibinfo {pages} {594} (\bibinfo {year} {2010})}\BibitemShut {NoStop}%
\bibitem [{\citenamefont {De~Greve}\ \emph {et~al.}(2013)\citenamefont
  {De~Greve}, \citenamefont {McMahon}, \citenamefont {Yu}, \citenamefont
  {Pelc}, \citenamefont {Jones}, \citenamefont {Natarajan}, \citenamefont
  {Kim}, \citenamefont {Abe}, \citenamefont {Maier}, \citenamefont {Schneider},
  \citenamefont {Kamp}, \citenamefont {H\"{o}fling}, \citenamefont {Hadfield},
  \citenamefont {Forchel}, \citenamefont {Fejer},\ and\ \citenamefont
  {Yamamoto}}]{deGreve2013complete}%
  \BibitemOpen
  \bibfield  {author} {\bibinfo {author} {\bibfnamefont {K.}~\bibnamefont
  {De~Greve}}, \bibinfo {author} {\bibfnamefont {P.~L.}\ \bibnamefont
  {McMahon}}, \bibinfo {author} {\bibfnamefont {L.}~\bibnamefont {Yu}},
  \bibinfo {author} {\bibfnamefont {J.~S.}\ \bibnamefont {Pelc}}, \bibinfo
  {author} {\bibfnamefont {C.}~\bibnamefont {Jones}}, \bibinfo {author}
  {\bibfnamefont {C.~M.}\ \bibnamefont {Natarajan}}, \bibinfo {author}
  {\bibfnamefont {N.~Y.}\ \bibnamefont {Kim}}, \bibinfo {author} {\bibfnamefont
  {E.}~\bibnamefont {Abe}}, \bibinfo {author} {\bibfnamefont {S.}~\bibnamefont
  {Maier}}, \bibinfo {author} {\bibfnamefont {C.}~\bibnamefont {Schneider}},
  \bibinfo {author} {\bibfnamefont {M.}~\bibnamefont {Kamp}}, \bibinfo {author}
  {\bibfnamefont {S.}~\bibnamefont {H\"{o}fling}}, \bibinfo {author}
  {\bibfnamefont {R.~H.}\ \bibnamefont {Hadfield}}, \bibinfo {author}
  {\bibfnamefont {A.}~\bibnamefont {Forchel}}, \bibinfo {author} {\bibfnamefont
  {M.~M.}\ \bibnamefont {Fejer}}, \ and\ \bibinfo {author} {\bibfnamefont
  {Y.}~\bibnamefont {Yamamoto}},\ }\href {\doibase 10.1038/ncomms3228}
  {\bibfield  {journal} {\bibinfo  {journal} {Nature Communications}\ }\textbf
  {\bibinfo {volume} {4}},\ \bibinfo {pages} {2228} (\bibinfo {year}
  {2013})}\BibitemShut {NoStop}%
\bibitem [{\citenamefont {Gao}\ \emph {et~al.}(2012)\citenamefont {Gao},
  \citenamefont {Fallahi}, \citenamefont {Togan}, \citenamefont
  {Miguel-Sanchez},\ and\ \citenamefont {Imamoglu}}]{gao2012observation}%
  \BibitemOpen
  \bibfield  {author} {\bibinfo {author} {\bibfnamefont {W.}~\bibnamefont
  {Gao}}, \bibinfo {author} {\bibfnamefont {P.}~\bibnamefont {Fallahi}},
  \bibinfo {author} {\bibfnamefont {E.}~\bibnamefont {Togan}}, \bibinfo
  {author} {\bibfnamefont {J.}~\bibnamefont {Miguel-Sanchez}}, \ and\ \bibinfo
  {author} {\bibfnamefont {A.}~\bibnamefont {Imamoglu}},\ }\href@noop {}
  {\bibfield  {journal} {\bibinfo  {journal} {Nature}\ }\textbf {\bibinfo
  {volume} {491}},\ \bibinfo {pages} {426} (\bibinfo {year}
  {2012})}\BibitemShut {NoStop}%
\bibitem [{\citenamefont {Schaibley}\ \emph {et~al.}(2013)\citenamefont
  {Schaibley}, \citenamefont {Burgers}, \citenamefont {McCracken},
  \citenamefont {Duan}, \citenamefont {Berman}, \citenamefont {Steel},
  \citenamefont {Bracker}, \citenamefont {Gammon},\ and\ \citenamefont
  {Sham}}]{schaibley2013demonstration}%
  \BibitemOpen
  \bibfield  {author} {\bibinfo {author} {\bibfnamefont {J.}~\bibnamefont
  {Schaibley}}, \bibinfo {author} {\bibfnamefont {A.}~\bibnamefont {Burgers}},
  \bibinfo {author} {\bibfnamefont {G.}~\bibnamefont {McCracken}}, \bibinfo
  {author} {\bibfnamefont {L.-M.}\ \bibnamefont {Duan}}, \bibinfo {author}
  {\bibfnamefont {P.}~\bibnamefont {Berman}}, \bibinfo {author} {\bibfnamefont
  {D.}~\bibnamefont {Steel}}, \bibinfo {author} {\bibfnamefont
  {A.}~\bibnamefont {Bracker}}, \bibinfo {author} {\bibfnamefont
  {D.}~\bibnamefont {Gammon}}, \ and\ \bibinfo {author} {\bibfnamefont
  {L.}~\bibnamefont {Sham}},\ }\href@noop {} {\bibfield  {journal} {\bibinfo
  {journal} {Physical Review Letters}\ }\textbf {\bibinfo {volume} {110}},\
  \bibinfo {pages} {167401} (\bibinfo {year} {2013})}\BibitemShut {NoStop}%
\bibitem [{\citenamefont {Press}\ \emph {et~al.}(2008)\citenamefont {Press},
  \citenamefont {Ladd}, \citenamefont {Zhang},\ and\ \citenamefont
  {Yamamoto}}]{press2008complete}%
  \BibitemOpen
  \bibfield  {author} {\bibinfo {author} {\bibfnamefont {D.}~\bibnamefont
  {Press}}, \bibinfo {author} {\bibfnamefont {T.~D.}\ \bibnamefont {Ladd}},
  \bibinfo {author} {\bibfnamefont {B.}~\bibnamefont {Zhang}}, \ and\ \bibinfo
  {author} {\bibfnamefont {Y.}~\bibnamefont {Yamamoto}},\ }\href@noop {}
  {\bibfield  {journal} {\bibinfo  {journal} {Nature}\ }\textbf {\bibinfo
  {volume} {456}},\ \bibinfo {pages} {218} (\bibinfo {year}
  {2008})}\BibitemShut {NoStop}%
\bibitem [{\citenamefont {Brunner}\ \emph {et~al.}(2009)\citenamefont
  {Brunner}, \citenamefont {Gerardot}, \citenamefont {Dalgarno}, \citenamefont
  {W{\"u}st}, \citenamefont {Karrai}, \citenamefont {Stoltz}, \citenamefont
  {Petroff},\ and\ \citenamefont {Warburton}}]{brunner2009coherent}%
  \BibitemOpen
  \bibfield  {author} {\bibinfo {author} {\bibfnamefont {D.}~\bibnamefont
  {Brunner}}, \bibinfo {author} {\bibfnamefont {B.~D.}\ \bibnamefont
  {Gerardot}}, \bibinfo {author} {\bibfnamefont {P.~A.}\ \bibnamefont
  {Dalgarno}}, \bibinfo {author} {\bibfnamefont {G.}~\bibnamefont {W{\"u}st}},
  \bibinfo {author} {\bibfnamefont {K.}~\bibnamefont {Karrai}}, \bibinfo
  {author} {\bibfnamefont {N.~G.}\ \bibnamefont {Stoltz}}, \bibinfo {author}
  {\bibfnamefont {P.~M.}\ \bibnamefont {Petroff}}, \ and\ \bibinfo {author}
  {\bibfnamefont {R.~J.}\ \bibnamefont {Warburton}},\ }\href@noop {} {\bibfield
   {journal} {\bibinfo  {journal} {Science}\ }\textbf {\bibinfo {volume}
  {325}},\ \bibinfo {pages} {70} (\bibinfo {year} {2009})}\BibitemShut
  {NoStop}%
\bibitem [{\citenamefont {Knill}\ \emph {et~al.}(2001)\citenamefont {Knill},
  \citenamefont {Laflamme},\ and\ \citenamefont {Milburn}}]{knill2001scheme}%
  \BibitemOpen
  \bibfield  {author} {\bibinfo {author} {\bibfnamefont {E.}~\bibnamefont
  {Knill}}, \bibinfo {author} {\bibfnamefont {R.}~\bibnamefont {Laflamme}}, \
  and\ \bibinfo {author} {\bibfnamefont {G.~J.}\ \bibnamefont {Milburn}},\
  }\href@noop {} {\bibfield  {journal} {\bibinfo  {journal} {Nature}\ }\textbf
  {\bibinfo {volume} {409}},\ \bibinfo {pages} {46} (\bibinfo {year}
  {2001})}\BibitemShut {NoStop}%
\bibitem [{\citenamefont {Kok}\ and\ \citenamefont
  {Lovett}(2010)}]{kok2010introduction}%
  \BibitemOpen
  \bibfield  {author} {\bibinfo {author} {\bibfnamefont {P.}~\bibnamefont
  {Kok}}\ and\ \bibinfo {author} {\bibfnamefont {B.~W.}\ \bibnamefont
  {Lovett}},\ }\href {http://www.worldcat.org/isbn/0521519144} {}\bibinfo
  {edition} {1st}\ ed.\ (\bibinfo  {publisher} {Cambridge University Press},\
  \bibinfo {year} {2010})\BibitemShut {NoStop}%
\bibitem [{\citenamefont {Kimble}(2008)}]{kimble2008quantum}%
  \BibitemOpen
  \bibfield  {author} {\bibinfo {author} {\bibfnamefont {H.}~\bibnamefont
  {Kimble}},\ }\href@noop {} {\bibfield  {journal} {\bibinfo  {journal}
  {Nature}\ }\textbf {\bibinfo {volume} {453}},\ \bibinfo {pages} {1023}
  (\bibinfo {year} {2008})}\BibitemShut {NoStop}%
\bibitem [{\citenamefont {Sangouard}\ \emph {et~al.}(2011)\citenamefont
  {Sangouard}, \citenamefont {Simon}, \citenamefont {de~Riedmatten},\ and\
  \citenamefont {Gisin}}]{RevModPhys.83.33}%
  \BibitemOpen
  \bibfield  {author} {\bibinfo {author} {\bibfnamefont {N.}~\bibnamefont
  {Sangouard}}, \bibinfo {author} {\bibfnamefont {C.}~\bibnamefont {Simon}},
  \bibinfo {author} {\bibfnamefont {H.}~\bibnamefont {de~Riedmatten}}, \ and\
  \bibinfo {author} {\bibfnamefont {N.}~\bibnamefont {Gisin}},\ }\href
  {\doibase 10.1103/RevModPhys.83.33} {\bibfield  {journal} {\bibinfo
  {journal} {Rev. Mod. Phys.}\ }\textbf {\bibinfo {volume} {83}},\ \bibinfo
  {pages} {33} (\bibinfo {year} {2011})}\BibitemShut {NoStop}%
\bibitem [{\citenamefont {Patel}\ \emph {et~al.}(2010)\citenamefont {Patel},
  \citenamefont {Bennett}, \citenamefont {Farrer}, \citenamefont {Nicoll},
  \citenamefont {Ritchie},\ and\ \citenamefont {Shields}}]{patel2010two}%
  \BibitemOpen
  \bibfield  {author} {\bibinfo {author} {\bibfnamefont {R.~B.}\ \bibnamefont
  {Patel}}, \bibinfo {author} {\bibfnamefont {A.~J.}\ \bibnamefont {Bennett}},
  \bibinfo {author} {\bibfnamefont {I.}~\bibnamefont {Farrer}}, \bibinfo
  {author} {\bibfnamefont {C.~A.}\ \bibnamefont {Nicoll}}, \bibinfo {author}
  {\bibfnamefont {D.~A.}\ \bibnamefont {Ritchie}}, \ and\ \bibinfo {author}
  {\bibfnamefont {A.~J.}\ \bibnamefont {Shields}},\ }\href {\doibase
  10.1038/nphoton.2010.161} {\bibfield  {journal} {\bibinfo  {journal} {Nature
  Photonics}\ }\textbf {\bibinfo {volume} {4}},\ \bibinfo {pages} {632}
  (\bibinfo {year} {2010})}\BibitemShut {NoStop}%
\bibitem [{\citenamefont {Flagg}\ \emph {et~al.}(2010)\citenamefont {Flagg},
  \citenamefont {Muller}, \citenamefont {Polyakov}, \citenamefont {Ling},
  \citenamefont {Migdall},\ and\ \citenamefont
  {Solomon}}]{PhysRevLett.104.137401}%
  \BibitemOpen
  \bibfield  {author} {\bibinfo {author} {\bibfnamefont {E.~B.}\ \bibnamefont
  {Flagg}}, \bibinfo {author} {\bibfnamefont {A.}~\bibnamefont {Muller}},
  \bibinfo {author} {\bibfnamefont {S.~V.}\ \bibnamefont {Polyakov}}, \bibinfo
  {author} {\bibfnamefont {A.}~\bibnamefont {Ling}}, \bibinfo {author}
  {\bibfnamefont {A.}~\bibnamefont {Migdall}}, \ and\ \bibinfo {author}
  {\bibfnamefont {G.~S.}\ \bibnamefont {Solomon}},\ }\href {\doibase
  10.1103/PhysRevLett.104.137401} {\bibfield  {journal} {\bibinfo  {journal}
  {Physical Review Letters}\ }\textbf {\bibinfo {volume} {104}},\ \bibinfo
  {pages} {137401} (\bibinfo {year} {2010})}\BibitemShut {NoStop}%
\bibitem [{\citenamefont {Strauf}\ \emph {et~al.}(2007)\citenamefont {Strauf},
  \citenamefont {Stoltz}, \citenamefont {Rakher}, \citenamefont {Coldren},
  \citenamefont {Petroff},\ and\ \citenamefont {Bouwmeester}}]{strauf2007high}%
  \BibitemOpen
  \bibfield  {author} {\bibinfo {author} {\bibfnamefont {S.}~\bibnamefont
  {Strauf}}, \bibinfo {author} {\bibfnamefont {N.~G.}\ \bibnamefont {Stoltz}},
  \bibinfo {author} {\bibfnamefont {M.~T.}\ \bibnamefont {Rakher}}, \bibinfo
  {author} {\bibfnamefont {L.~A.}\ \bibnamefont {Coldren}}, \bibinfo {author}
  {\bibfnamefont {P.~M.}\ \bibnamefont {Petroff}}, \ and\ \bibinfo {author}
  {\bibfnamefont {D.}~\bibnamefont {Bouwmeester}},\ }\href@noop {} {\bibfield
  {journal} {\bibinfo  {journal} {Nature Photonics}\ }\textbf {\bibinfo
  {volume} {1}},\ \bibinfo {pages} {704} (\bibinfo {year} {2007})}\BibitemShut
  {NoStop}%
\bibitem [{\citenamefont {Gazzano}\ \emph {et~al.}(2013)\citenamefont
  {Gazzano}, \citenamefont {de~Vasconcellos}, \citenamefont {Arnold},
  \citenamefont {Nowak}, \citenamefont {Galopin}, \citenamefont {Sagnes},
  \citenamefont {Lanco}, \citenamefont {Lema{\^\i}tre},\ and\ \citenamefont
  {Senellart}}]{gazzano2013bright}%
  \BibitemOpen
  \bibfield  {author} {\bibinfo {author} {\bibfnamefont {O.}~\bibnamefont
  {Gazzano}}, \bibinfo {author} {\bibfnamefont {S.~M.}\ \bibnamefont
  {de~Vasconcellos}}, \bibinfo {author} {\bibfnamefont {C.}~\bibnamefont
  {Arnold}}, \bibinfo {author} {\bibfnamefont {A.}~\bibnamefont {Nowak}},
  \bibinfo {author} {\bibfnamefont {E.}~\bibnamefont {Galopin}}, \bibinfo
  {author} {\bibfnamefont {I.}~\bibnamefont {Sagnes}}, \bibinfo {author}
  {\bibfnamefont {L.}~\bibnamefont {Lanco}}, \bibinfo {author} {\bibfnamefont
  {A.}~\bibnamefont {Lema{\^\i}tre}}, \ and\ \bibinfo {author} {\bibfnamefont
  {P.}~\bibnamefont {Senellart}},\ }\href@noop {} {\bibfield  {journal}
  {\bibinfo  {journal} {Nature Communications}\ }\textbf {\bibinfo {volume}
  {4}},\ \bibinfo {pages} {1425} (\bibinfo {year} {2013})}\BibitemShut
  {NoStop}%
\bibitem [{\citenamefont {Madsen}\ \emph {et~al.}(2014)\citenamefont {Madsen},
  \citenamefont {Ates}, \citenamefont {Liu}, \citenamefont {Javadi},
  \citenamefont {Albrecht}, \citenamefont {Yeo}, \citenamefont {Stobbe},\ and\
  \citenamefont {Lodahl}}]{madsen2014efficient}%
  \BibitemOpen
  \bibfield  {author} {\bibinfo {author} {\bibfnamefont {K.~H.}\ \bibnamefont
  {Madsen}}, \bibinfo {author} {\bibfnamefont {S.}~\bibnamefont {Ates}},
  \bibinfo {author} {\bibfnamefont {J.}~\bibnamefont {Liu}}, \bibinfo {author}
  {\bibfnamefont {A.}~\bibnamefont {Javadi}}, \bibinfo {author} {\bibfnamefont
  {S.~M.}\ \bibnamefont {Albrecht}}, \bibinfo {author} {\bibfnamefont
  {I.}~\bibnamefont {Yeo}}, \bibinfo {author} {\bibfnamefont {S.}~\bibnamefont
  {Stobbe}}, \ and\ \bibinfo {author} {\bibfnamefont {P.}~\bibnamefont
  {Lodahl}},\ }\href {\doibase 10.1103/PhysRevB.90.155303} {\bibfield
  {journal} {\bibinfo  {journal} {Phys. Rev. B}\ }\textbf {\bibinfo {volume}
  {90}},\ \bibinfo {pages} {155303} (\bibinfo {year} {2014})}\BibitemShut
  {NoStop}%
\bibitem [{\citenamefont {Davan{\c{c}}o}\ \emph {et~al.}(2011)\citenamefont
  {Davan{\c{c}}o}, \citenamefont {Rakher}, \citenamefont {Schuh}, \citenamefont
  {Badolato},\ and\ \citenamefont {Srinivasan}}]{davancco2011circular}%
  \BibitemOpen
  \bibfield  {author} {\bibinfo {author} {\bibfnamefont {M.}~\bibnamefont
  {Davan{\c{c}}o}}, \bibinfo {author} {\bibfnamefont {M.}~\bibnamefont
  {Rakher}}, \bibinfo {author} {\bibfnamefont {D.}~\bibnamefont {Schuh}},
  \bibinfo {author} {\bibfnamefont {A.}~\bibnamefont {Badolato}}, \ and\
  \bibinfo {author} {\bibfnamefont {K.}~\bibnamefont {Srinivasan}},\
  }\href@noop {} {\bibfield  {journal} {\bibinfo  {journal} {Applied Physics
  Letters}\ }\textbf {\bibinfo {volume} {99}},\ \bibinfo {pages} {041102}
  (\bibinfo {year} {2011})}\BibitemShut {NoStop}%
\bibitem [{\citenamefont {Trotta}\ \emph {et~al.}(2012)\citenamefont {Trotta},
  \citenamefont {Atkinson}, \citenamefont {Plumhof}, \citenamefont {Zallo},
  \citenamefont {Rezaev}, \citenamefont {Kumar}, \citenamefont {Baunack},
  \citenamefont {Schroeter}, \citenamefont {Rastelli},\ and\ \citenamefont
  {Schmidt}}]{trotta2012nanomembrane}%
  \BibitemOpen
  \bibfield  {author} {\bibinfo {author} {\bibfnamefont {R.}~\bibnamefont
  {Trotta}}, \bibinfo {author} {\bibfnamefont {P.}~\bibnamefont {Atkinson}},
  \bibinfo {author} {\bibfnamefont {J.}~\bibnamefont {Plumhof}}, \bibinfo
  {author} {\bibfnamefont {E.}~\bibnamefont {Zallo}}, \bibinfo {author}
  {\bibfnamefont {R.}~\bibnamefont {Rezaev}}, \bibinfo {author} {\bibfnamefont
  {S.}~\bibnamefont {Kumar}}, \bibinfo {author} {\bibfnamefont
  {S.}~\bibnamefont {Baunack}}, \bibinfo {author} {\bibfnamefont
  {J.}~\bibnamefont {Schroeter}}, \bibinfo {author} {\bibfnamefont
  {A.}~\bibnamefont {Rastelli}}, \ and\ \bibinfo {author} {\bibfnamefont
  {O.}~\bibnamefont {Schmidt}},\ }\href@noop {} {\bibfield  {journal} {\bibinfo
   {journal} {Advanced Materials}\ }\textbf {\bibinfo {volume} {24}},\ \bibinfo
  {pages} {2668} (\bibinfo {year} {2012})}\BibitemShut {NoStop}%
\bibitem [{\citenamefont {Ma}\ \emph {et~al.}(2014)\citenamefont {Ma},
  \citenamefont {Kremer},\ and\ \citenamefont {Gerardot}}]{ma2014efficient}%
  \BibitemOpen
  \bibfield  {author} {\bibinfo {author} {\bibfnamefont {Y.}~\bibnamefont
  {Ma}}, \bibinfo {author} {\bibfnamefont {P.~E.}\ \bibnamefont {Kremer}}, \
  and\ \bibinfo {author} {\bibfnamefont {B.~D.}\ \bibnamefont {Gerardot}},\
  }\href@noop {} {\bibfield  {journal} {\bibinfo  {journal} {Journal of Applied
  Physics}\ }\textbf {\bibinfo {volume} {115}},\ \bibinfo {pages} {023106}
  (\bibinfo {year} {2014})}\BibitemShut {NoStop}%
\bibitem [{\citenamefont {Claudon}\ \emph {et~al.}(2010)\citenamefont
  {Claudon}, \citenamefont {Bleuse}, \citenamefont {Malik}, \citenamefont
  {Bazin}, \citenamefont {Jaffrennou}, \citenamefont {Gregersen}, \citenamefont
  {Sauvan}, \citenamefont {Lalanne},\ and\ \citenamefont
  {G{\'e}rard}}]{claudon2010highly}%
  \BibitemOpen
  \bibfield  {author} {\bibinfo {author} {\bibfnamefont {J.}~\bibnamefont
  {Claudon}}, \bibinfo {author} {\bibfnamefont {J.}~\bibnamefont {Bleuse}},
  \bibinfo {author} {\bibfnamefont {N.~S.}\ \bibnamefont {Malik}}, \bibinfo
  {author} {\bibfnamefont {M.}~\bibnamefont {Bazin}}, \bibinfo {author}
  {\bibfnamefont {P.}~\bibnamefont {Jaffrennou}}, \bibinfo {author}
  {\bibfnamefont {N.}~\bibnamefont {Gregersen}}, \bibinfo {author}
  {\bibfnamefont {C.}~\bibnamefont {Sauvan}}, \bibinfo {author} {\bibfnamefont
  {P.}~\bibnamefont {Lalanne}}, \ and\ \bibinfo {author} {\bibfnamefont
  {J.-M.}\ \bibnamefont {G{\'e}rard}},\ }\href@noop {} {\bibfield  {journal}
  {\bibinfo  {journal} {Nature Photonics}\ }\textbf {\bibinfo {volume} {4}},\
  \bibinfo {pages} {174} (\bibinfo {year} {2010})}\BibitemShut {NoStop}%
\bibitem [{\citenamefont {Reimer}\ \emph {et~al.}(2012)\citenamefont {Reimer},
  \citenamefont {Bulgarini}, \citenamefont {Akopian}, \citenamefont {Hocevar},
  \citenamefont {Bavinck}, \citenamefont {Verheijen}, \citenamefont {Bakkers},
  \citenamefont {Kouwenhoven},\ and\ \citenamefont
  {Zwiller}}]{reimer2012bright}%
  \BibitemOpen
  \bibfield  {author} {\bibinfo {author} {\bibfnamefont {M.~E.}\ \bibnamefont
  {Reimer}}, \bibinfo {author} {\bibfnamefont {G.}~\bibnamefont {Bulgarini}},
  \bibinfo {author} {\bibfnamefont {N.}~\bibnamefont {Akopian}}, \bibinfo
  {author} {\bibfnamefont {M.}~\bibnamefont {Hocevar}}, \bibinfo {author}
  {\bibfnamefont {M.~B.}\ \bibnamefont {Bavinck}}, \bibinfo {author}
  {\bibfnamefont {M.~A.}\ \bibnamefont {Verheijen}}, \bibinfo {author}
  {\bibfnamefont {E.~P.}\ \bibnamefont {Bakkers}}, \bibinfo {author}
  {\bibfnamefont {L.~P.}\ \bibnamefont {Kouwenhoven}}, \ and\ \bibinfo {author}
  {\bibfnamefont {V.}~\bibnamefont {Zwiller}},\ }\href@noop {} {\bibfield
  {journal} {\bibinfo  {journal} {Nature Communications}\ }\textbf {\bibinfo
  {volume} {3}},\ \bibinfo {pages} {737} (\bibinfo {year} {2012})}\BibitemShut
  {NoStop}%
\bibitem [{\citenamefont {Gregersen}\ \emph {et~al.}(2008)\citenamefont
  {Gregersen}, \citenamefont {Nielsen}, \citenamefont {Claudon}, \citenamefont
  {G{\'e}rard},\ and\ \citenamefont {M{\o}rk}}]{gregersen2008controlling}%
  \BibitemOpen
  \bibfield  {author} {\bibinfo {author} {\bibfnamefont {N.}~\bibnamefont
  {Gregersen}}, \bibinfo {author} {\bibfnamefont {T.~R.}\ \bibnamefont
  {Nielsen}}, \bibinfo {author} {\bibfnamefont {J.}~\bibnamefont {Claudon}},
  \bibinfo {author} {\bibfnamefont {J.-M.}\ \bibnamefont {G{\'e}rard}}, \ and\
  \bibinfo {author} {\bibfnamefont {J.}~\bibnamefont {M{\o}rk}},\ }\href@noop
  {} {\bibfield  {journal} {\bibinfo  {journal} {Optics Letters}\ }\textbf
  {\bibinfo {volume} {33}},\ \bibinfo {pages} {1693} (\bibinfo {year}
  {2008})}\BibitemShut {NoStop}%
\bibitem [{\citenamefont {Friedler}\ \emph {et~al.}(2009)\citenamefont
  {Friedler}, \citenamefont {Sauvan}, \citenamefont {Hugonin}, \citenamefont
  {Lalanne}, \citenamefont {Claudon},\ and\ \citenamefont
  {G{\'e}rard}}]{friedler2009solid}%
  \BibitemOpen
  \bibfield  {author} {\bibinfo {author} {\bibfnamefont {I.}~\bibnamefont
  {Friedler}}, \bibinfo {author} {\bibfnamefont {C.}~\bibnamefont {Sauvan}},
  \bibinfo {author} {\bibfnamefont {J.-P.}\ \bibnamefont {Hugonin}}, \bibinfo
  {author} {\bibfnamefont {P.}~\bibnamefont {Lalanne}}, \bibinfo {author}
  {\bibfnamefont {J.}~\bibnamefont {Claudon}}, \ and\ \bibinfo {author}
  {\bibfnamefont {J.-M.}\ \bibnamefont {G{\'e}rard}},\ }\href@noop {}
  {\bibfield  {journal} {\bibinfo  {journal} {Optics Express}\ }\textbf
  {\bibinfo {volume} {17}},\ \bibinfo {pages} {2095} (\bibinfo {year}
  {2009})}\BibitemShut {NoStop}%
\bibitem [{\citenamefont {Claudon}\ \emph {et~al.}(2013)\citenamefont
  {Claudon}, \citenamefont {Gregersen}, \citenamefont {Lalanne},\ and\
  \citenamefont {G{\'e}rard}}]{claudon2013harnessing}%
  \BibitemOpen
  \bibfield  {author} {\bibinfo {author} {\bibfnamefont {J.}~\bibnamefont
  {Claudon}}, \bibinfo {author} {\bibfnamefont {N.}~\bibnamefont {Gregersen}},
  \bibinfo {author} {\bibfnamefont {P.}~\bibnamefont {Lalanne}}, \ and\
  \bibinfo {author} {\bibfnamefont {J.-M.}\ \bibnamefont {G{\'e}rard}},\
  }\href@noop {} {\bibfield  {journal} {\bibinfo  {journal} {ChemPhysChem}\
  }\textbf {\bibinfo {volume} {14}},\ \bibinfo {pages} {2393} (\bibinfo {year}
  {2013})}\BibitemShut {NoStop}%
\bibitem [{\citenamefont {Finley}\ \emph {et~al.}(2001)\citenamefont {Finley},
  \citenamefont {Fry}, \citenamefont {Ashmore}, \citenamefont {Lema\^{i}tre},
  \citenamefont {Tartakovskii}, \citenamefont {Oulton}, \citenamefont
  {Mowbray}, \citenamefont {Skolnick}, \citenamefont {Hopkinson}, \citenamefont
  {Buckle},\ and\ \citenamefont {Maksym}}]{finley2001observation}%
  \BibitemOpen
  \bibfield  {author} {\bibinfo {author} {\bibfnamefont {J.~J.}\ \bibnamefont
  {Finley}}, \bibinfo {author} {\bibfnamefont {P.~W.}\ \bibnamefont {Fry}},
  \bibinfo {author} {\bibfnamefont {A.~D.}\ \bibnamefont {Ashmore}}, \bibinfo
  {author} {\bibfnamefont {A.}~\bibnamefont {Lema\^{i}tre}}, \bibinfo {author}
  {\bibfnamefont {A.~I.}\ \bibnamefont {Tartakovskii}}, \bibinfo {author}
  {\bibfnamefont {R.}~\bibnamefont {Oulton}}, \bibinfo {author} {\bibfnamefont
  {D.~J.}\ \bibnamefont {Mowbray}}, \bibinfo {author} {\bibfnamefont {M.~S.}\
  \bibnamefont {Skolnick}}, \bibinfo {author} {\bibfnamefont {M.}~\bibnamefont
  {Hopkinson}}, \bibinfo {author} {\bibfnamefont {P.~D.}\ \bibnamefont
  {Buckle}}, \ and\ \bibinfo {author} {\bibfnamefont {P.~A.}\ \bibnamefont
  {Maksym}},\ }\href@noop {} {\bibfield  {journal} {\bibinfo  {journal}
  {Physical Review B}\ }\textbf {\bibinfo {volume} {63}},\ \bibinfo {pages}
  {161305} (\bibinfo {year} {2001})}\BibitemShut {NoStop}%
\bibitem [{\citenamefont {H{\"o}gele}\ \emph {et~al.}(2004)\citenamefont
  {H{\"o}gele}, \citenamefont {Seidl}, \citenamefont {Kroner}, \citenamefont
  {Karrai}, \citenamefont {Warburton}, \citenamefont {Gerardot},\ and\
  \citenamefont {Petroff}}]{hogele2004voltage}%
  \BibitemOpen
  \bibfield  {author} {\bibinfo {author} {\bibfnamefont {A.}~\bibnamefont
  {H{\"o}gele}}, \bibinfo {author} {\bibfnamefont {S.}~\bibnamefont {Seidl}},
  \bibinfo {author} {\bibfnamefont {M.}~\bibnamefont {Kroner}}, \bibinfo
  {author} {\bibfnamefont {K.}~\bibnamefont {Karrai}}, \bibinfo {author}
  {\bibfnamefont {R.~J.}\ \bibnamefont {Warburton}}, \bibinfo {author}
  {\bibfnamefont {B.~D.}\ \bibnamefont {Gerardot}}, \ and\ \bibinfo {author}
  {\bibfnamefont {P.~M.}\ \bibnamefont {Petroff}},\ }\href@noop {} {\bibfield
  {journal} {\bibinfo  {journal} {Physical Review Letters}\ }\textbf {\bibinfo
  {volume} {93}},\ \bibinfo {pages} {217401} (\bibinfo {year}
  {2004})}\BibitemShut {NoStop}%
\bibitem [{\citenamefont {Gerardot}\ \emph {et~al.}(2007)\citenamefont
  {Gerardot}, \citenamefont {Seidl}, \citenamefont {Dalgarno}, \citenamefont
  {Warburton}, \citenamefont {Granados}, \citenamefont {Garcia}, \citenamefont
  {Kowalik}, \citenamefont {Krebs}, \citenamefont {Karrai}, \citenamefont
  {Badolato},\ and\ \citenamefont {Petroff}}]{gerardot2007manipulating}%
  \BibitemOpen
  \bibfield  {author} {\bibinfo {author} {\bibfnamefont {B.~D.}\ \bibnamefont
  {Gerardot}}, \bibinfo {author} {\bibfnamefont {S.}~\bibnamefont {Seidl}},
  \bibinfo {author} {\bibfnamefont {P.~A.}\ \bibnamefont {Dalgarno}}, \bibinfo
  {author} {\bibfnamefont {R.~J.}\ \bibnamefont {Warburton}}, \bibinfo {author}
  {\bibfnamefont {D.}~\bibnamefont {Granados}}, \bibinfo {author}
  {\bibfnamefont {J.~M.}\ \bibnamefont {Garcia}}, \bibinfo {author}
  {\bibfnamefont {K.}~\bibnamefont {Kowalik}}, \bibinfo {author} {\bibfnamefont
  {O.}~\bibnamefont {Krebs}}, \bibinfo {author} {\bibfnamefont
  {K.}~\bibnamefont {Karrai}}, \bibinfo {author} {\bibfnamefont
  {A.}~\bibnamefont {Badolato}}, \ and\ \bibinfo {author} {\bibfnamefont
  {P.~M.}\ \bibnamefont {Petroff}},\ }\href@noop {} {\bibfield  {journal}
  {\bibinfo  {journal} {Applied Physics Letters}\ }\textbf {\bibinfo {volume}
  {90}},\ \bibinfo {pages} {041101} (\bibinfo {year} {2007})}\BibitemShut
  {NoStop}%
\bibitem [{\citenamefont {Vogel}\ \emph {et~al.}(2007)\citenamefont {Vogel},
  \citenamefont {Ulrich}, \citenamefont {Hafenbrak}, \citenamefont {Michler},
  \citenamefont {Wang}, \citenamefont {Rastelli},\ and\ \citenamefont
  {Schmidt}}]{vogel2007influence}%
  \BibitemOpen
  \bibfield  {author} {\bibinfo {author} {\bibfnamefont {M.}~\bibnamefont
  {Vogel}}, \bibinfo {author} {\bibfnamefont {S.}~\bibnamefont {Ulrich}},
  \bibinfo {author} {\bibfnamefont {R.}~\bibnamefont {Hafenbrak}}, \bibinfo
  {author} {\bibfnamefont {P.}~\bibnamefont {Michler}}, \bibinfo {author}
  {\bibfnamefont {L.}~\bibnamefont {Wang}}, \bibinfo {author} {\bibfnamefont
  {A.}~\bibnamefont {Rastelli}}, \ and\ \bibinfo {author} {\bibfnamefont
  {O.}~\bibnamefont {Schmidt}},\ }\href@noop {} {\bibfield  {journal} {\bibinfo
   {journal} {Applied Physics Letters}\ }\textbf {\bibinfo {volume} {91}},\
  \bibinfo {pages} {051904} (\bibinfo {year} {2007})}\BibitemShut {NoStop}%
\bibitem [{\citenamefont {Seidl}\ \emph {et~al.}(2006)\citenamefont {Seidl},
  \citenamefont {Kroner}, \citenamefont {H{\"o}gele}, \citenamefont {Karrai},
  \citenamefont {Warburton}, \citenamefont {Badolato},\ and\ \citenamefont
  {Petroff}}]{seidl2006effect}%
  \BibitemOpen
  \bibfield  {author} {\bibinfo {author} {\bibfnamefont {S.}~\bibnamefont
  {Seidl}}, \bibinfo {author} {\bibfnamefont {M.}~\bibnamefont {Kroner}},
  \bibinfo {author} {\bibfnamefont {A.}~\bibnamefont {H{\"o}gele}}, \bibinfo
  {author} {\bibfnamefont {K.}~\bibnamefont {Karrai}}, \bibinfo {author}
  {\bibfnamefont {R.~J.}\ \bibnamefont {Warburton}}, \bibinfo {author}
  {\bibfnamefont {A.}~\bibnamefont {Badolato}}, \ and\ \bibinfo {author}
  {\bibfnamefont {P.~M.}\ \bibnamefont {Petroff}},\ }\href@noop {} {\bibfield
  {journal} {\bibinfo  {journal} {Applied Physics Letters}\ }\textbf {\bibinfo
  {volume} {88}},\ \bibinfo {pages} {203113} (\bibinfo {year}
  {2006})}\BibitemShut {NoStop}%
\bibitem [{\citenamefont {Kuklewicz}\ \emph {et~al.}(2012)\citenamefont
  {Kuklewicz}, \citenamefont {Malein}, \citenamefont {Petroff},\ and\
  \citenamefont {Gerardot}}]{kuklewicz2012electro}%
  \BibitemOpen
  \bibfield  {author} {\bibinfo {author} {\bibfnamefont {C.~E.}\ \bibnamefont
  {Kuklewicz}}, \bibinfo {author} {\bibfnamefont {R.~N.}\ \bibnamefont
  {Malein}}, \bibinfo {author} {\bibfnamefont {P.~M.}\ \bibnamefont {Petroff}},
  \ and\ \bibinfo {author} {\bibfnamefont {B.~D.}\ \bibnamefont {Gerardot}},\
  }\href@noop {} {\bibfield  {journal} {\bibinfo  {journal} {Nano Letters}\
  }\textbf {\bibinfo {volume} {12}},\ \bibinfo {pages} {3761} (\bibinfo {year}
  {2012})}\BibitemShut {NoStop}%
\bibitem [{\citenamefont {Sapienza}\ \emph {et~al.}(2013)\citenamefont
  {Sapienza}, \citenamefont {Malein}, \citenamefont {Kuklewicz}, \citenamefont
  {Kremer}, \citenamefont {Srinivasan}, \citenamefont {Griffiths},
  \citenamefont {Clarke}, \citenamefont {Gong}, \citenamefont {Warburton},\
  and\ \citenamefont {Gerardot}}]{sapienza2013exciton}%
  \BibitemOpen
  \bibfield  {author} {\bibinfo {author} {\bibfnamefont {L.}~\bibnamefont
  {Sapienza}}, \bibinfo {author} {\bibfnamefont {R.~N.}\ \bibnamefont
  {Malein}}, \bibinfo {author} {\bibfnamefont {C.~E.}\ \bibnamefont
  {Kuklewicz}}, \bibinfo {author} {\bibfnamefont {P.~E.}\ \bibnamefont
  {Kremer}}, \bibinfo {author} {\bibfnamefont {K.}~\bibnamefont {Srinivasan}},
  \bibinfo {author} {\bibfnamefont {A.}~\bibnamefont {Griffiths}}, \bibinfo
  {author} {\bibfnamefont {E.}~\bibnamefont {Clarke}}, \bibinfo {author}
  {\bibfnamefont {M.}~\bibnamefont {Gong}}, \bibinfo {author} {\bibfnamefont
  {R.~J.}\ \bibnamefont {Warburton}}, \ and\ \bibinfo {author} {\bibfnamefont
  {B.~D.}\ \bibnamefont {Gerardot}},\ }\href@noop {} {\bibfield  {journal}
  {\bibinfo  {journal} {Physical Review B}\ }\textbf {\bibinfo {volume} {88}},\
  \bibinfo {pages} {155330} (\bibinfo {year} {2013})}\BibitemShut {NoStop}%
\bibitem [{\citenamefont {Gregersen}\ \emph {et~al.}(2010)\citenamefont
  {Gregersen}, \citenamefont {Nielsen}, \citenamefont {M{\o}rk}, \citenamefont
  {Claudon},\ and\ \citenamefont {G{\'e}rard}}]{gregersen2010designs}%
  \BibitemOpen
  \bibfield  {author} {\bibinfo {author} {\bibfnamefont {N.}~\bibnamefont
  {Gregersen}}, \bibinfo {author} {\bibfnamefont {T.~R.}\ \bibnamefont
  {Nielsen}}, \bibinfo {author} {\bibfnamefont {J.}~\bibnamefont {M{\o}rk}},
  \bibinfo {author} {\bibfnamefont {J.}~\bibnamefont {Claudon}}, \ and\
  \bibinfo {author} {\bibfnamefont {J.-M.}\ \bibnamefont {G{\'e}rard}},\
  }\href@noop {} {\bibfield  {journal} {\bibinfo  {journal} {Optics Express}\
  }\textbf {\bibinfo {volume} {18}},\ \bibinfo {pages} {21204} (\bibinfo {year}
  {2010})}\BibitemShut {NoStop}%
\bibitem [{\citenamefont {Yeo}\ \emph {et~al.}(2013)\citenamefont {Yeo},
  \citenamefont {de~Assis}, \citenamefont {Gloppe}, \citenamefont
  {Dupont-Ferrier}, \citenamefont {Verlot}, \citenamefont {Malik},
  \citenamefont {Dupuy}, \citenamefont {Claudon}, \citenamefont {G{\'e}rard},
  \citenamefont {Auff{\`e}ves}, \citenamefont {Nogues}, \citenamefont
  {Seidelin}, \citenamefont {Poizat}, \citenamefont {Arcizet},\ and\
  \citenamefont {Richard}}]{yeo2013strain}%
  \BibitemOpen
  \bibfield  {author} {\bibinfo {author} {\bibfnamefont {I.}~\bibnamefont
  {Yeo}}, \bibinfo {author} {\bibfnamefont {P.-L.}\ \bibnamefont {de~Assis}},
  \bibinfo {author} {\bibfnamefont {A.}~\bibnamefont {Gloppe}}, \bibinfo
  {author} {\bibfnamefont {E.}~\bibnamefont {Dupont-Ferrier}}, \bibinfo
  {author} {\bibfnamefont {P.}~\bibnamefont {Verlot}}, \bibinfo {author}
  {\bibfnamefont {N.~S.}\ \bibnamefont {Malik}}, \bibinfo {author}
  {\bibfnamefont {E.}~\bibnamefont {Dupuy}}, \bibinfo {author} {\bibfnamefont
  {J.}~\bibnamefont {Claudon}}, \bibinfo {author} {\bibfnamefont {J.-M.}\
  \bibnamefont {G{\'e}rard}}, \bibinfo {author} {\bibfnamefont
  {A.}~\bibnamefont {Auff{\`e}ves}}, \bibinfo {author} {\bibfnamefont
  {G.}~\bibnamefont {Nogues}}, \bibinfo {author} {\bibfnamefont
  {S.}~\bibnamefont {Seidelin}}, \bibinfo {author} {\bibfnamefont {J.-P.}\
  \bibnamefont {Poizat}}, \bibinfo {author} {\bibfnamefont {O.}~\bibnamefont
  {Arcizet}}, \ and\ \bibinfo {author} {\bibfnamefont {M.}~\bibnamefont
  {Richard}},\ }\href@noop {} {\bibfield  {journal} {\bibinfo  {journal}
  {Nature Nanotechnology}\ } (\bibinfo {year} {2013})}\BibitemShut {NoStop}%
\bibitem [{\citenamefont {Montinaro}\ \emph {et~al.}(2014)\citenamefont
  {Montinaro}, \citenamefont {Wüst}, \citenamefont {Munsch}, \citenamefont
  {Fontana}, \citenamefont {Russo-Averchi}, \citenamefont {Heiss},
  \citenamefont {Fontcuberta~i Morral}, \citenamefont {Warburton},\ and\
  \citenamefont {Poggio}}]{montinaro2014quantum}%
  \BibitemOpen
  \bibfield  {author} {\bibinfo {author} {\bibfnamefont {M.}~\bibnamefont
  {Montinaro}}, \bibinfo {author} {\bibfnamefont {G.}~\bibnamefont {Wüst}},
  \bibinfo {author} {\bibfnamefont {M.}~\bibnamefont {Munsch}}, \bibinfo
  {author} {\bibfnamefont {Y.}~\bibnamefont {Fontana}}, \bibinfo {author}
  {\bibfnamefont {E.}~\bibnamefont {Russo-Averchi}}, \bibinfo {author}
  {\bibfnamefont {M.}~\bibnamefont {Heiss}}, \bibinfo {author} {\bibfnamefont
  {A.}~\bibnamefont {Fontcuberta~i Morral}}, \bibinfo {author} {\bibfnamefont
  {R.~J.}\ \bibnamefont {Warburton}}, \ and\ \bibinfo {author} {\bibfnamefont
  {M.}~\bibnamefont {Poggio}},\ }\href {\doibase 10.1021/nl501413t} {\bibfield
  {journal} {\bibinfo  {journal} {Nano Letters}\ }\textbf {\bibinfo {volume}
  {14}},\ \bibinfo {pages} {4454} (\bibinfo {year} {2014})}\BibitemShut
  {NoStop}%
\bibitem [{\citenamefont {Wei{\ss}}\ \emph {et~al.}(2014)\citenamefont
  {Wei{\ss}}, \citenamefont {Kinzel}, \citenamefont {Sch�lein}, \citenamefont
  {Heigl}, \citenamefont {Rudolph}, \citenamefont {Mork�tter}, \citenamefont
  {D�blinger}, \citenamefont {Bichler}, \citenamefont {Abstreiter},
  \citenamefont {Finley} \emph {et~al.}}]{weiss2014dynamic}%
  \BibitemOpen
  \bibfield  {author} {\bibinfo {author} {\bibfnamefont {M.}~\bibnamefont
  {Wei{\ss}}}, \bibinfo {author} {\bibfnamefont {J.~B.}\ \bibnamefont
  {Kinzel}}, \bibinfo {author} {\bibfnamefont {F.~J.}\ \bibnamefont
  {Sch�lein}}, \bibinfo {author} {\bibfnamefont {M.}~\bibnamefont {Heigl}},
  \bibinfo {author} {\bibfnamefont {D.}~\bibnamefont {Rudolph}}, \bibinfo
  {author} {\bibfnamefont {S.}~\bibnamefont {Mork�tter}}, \bibinfo {author}
  {\bibfnamefont {M.}~\bibnamefont {D�blinger}}, \bibinfo {author}
  {\bibfnamefont {M.}~\bibnamefont {Bichler}}, \bibinfo {author} {\bibfnamefont
  {G.}~\bibnamefont {Abstreiter}}, \bibinfo {author} {\bibfnamefont {J.~J.}\
  \bibnamefont {Finley}},  \emph {et~al.},\ }\href@noop {} {\bibfield
  {journal} {\bibinfo  {journal} {Nano letters}\ }\textbf {\bibinfo {volume}
  {14}},\ \bibinfo {pages} {2256} (\bibinfo {year} {2014})}\BibitemShut
  {NoStop}%
\bibitem [{\citenamefont {Bouwes~Bavinck}\ \emph {et~al.}(2012)\citenamefont
  {Bouwes~Bavinck}, \citenamefont {Zielinski}, \citenamefont {Witek},
  \citenamefont {Zehender}, \citenamefont {Bakkers},\ and\ \citenamefont
  {Zwiller}}]{bouwes2012controlling}%
  \BibitemOpen
  \bibfield  {author} {\bibinfo {author} {\bibfnamefont {M.}~\bibnamefont
  {Bouwes~Bavinck}}, \bibinfo {author} {\bibfnamefont {M.}~\bibnamefont
  {Zielinski}}, \bibinfo {author} {\bibfnamefont {B.~J.}\ \bibnamefont
  {Witek}}, \bibinfo {author} {\bibfnamefont {T.}~\bibnamefont {Zehender}},
  \bibinfo {author} {\bibfnamefont {E.~P.}\ \bibnamefont {Bakkers}}, \ and\
  \bibinfo {author} {\bibfnamefont {V.}~\bibnamefont {Zwiller}},\ }\href@noop
  {} {\bibfield  {journal} {\bibinfo  {journal} {Nano Letters}\ }\textbf
  {\bibinfo {volume} {12}},\ \bibinfo {pages} {6206} (\bibinfo {year}
  {2012})}\BibitemShut {NoStop}%
\bibitem [{\citenamefont {Houel}\ \emph {et~al.}(2012)\citenamefont {Houel},
  \citenamefont {Kuhlmann}, \citenamefont {Greuter}, \citenamefont {Xue},
  \citenamefont {Poggio}, \citenamefont {Gerardot}, \citenamefont {Dalgarno},
  \citenamefont {Badolato}, \citenamefont {Petroff}, \citenamefont {Ludwig},
  \citenamefont {Reuter}, \citenamefont {Wieck},\ and\ \citenamefont
  {Warburton}}]{PhysRevLett.108.107401}%
  \BibitemOpen
  \bibfield  {author} {\bibinfo {author} {\bibfnamefont {J.}~\bibnamefont
  {Houel}}, \bibinfo {author} {\bibfnamefont {A.~V.}\ \bibnamefont {Kuhlmann}},
  \bibinfo {author} {\bibfnamefont {L.}~\bibnamefont {Greuter}}, \bibinfo
  {author} {\bibfnamefont {F.}~\bibnamefont {Xue}}, \bibinfo {author}
  {\bibfnamefont {M.}~\bibnamefont {Poggio}}, \bibinfo {author} {\bibfnamefont
  {B.~D.}\ \bibnamefont {Gerardot}}, \bibinfo {author} {\bibfnamefont {P.~A.}\
  \bibnamefont {Dalgarno}}, \bibinfo {author} {\bibfnamefont {A.}~\bibnamefont
  {Badolato}}, \bibinfo {author} {\bibfnamefont {P.~M.}\ \bibnamefont
  {Petroff}}, \bibinfo {author} {\bibfnamefont {A.}~\bibnamefont {Ludwig}},
  \bibinfo {author} {\bibfnamefont {D.}~\bibnamefont {Reuter}}, \bibinfo
  {author} {\bibfnamefont {A.~D.}\ \bibnamefont {Wieck}}, \ and\ \bibinfo
  {author} {\bibfnamefont {R.~J.}\ \bibnamefont {Warburton}},\ }\href {\doibase
  10.1103/PhysRevLett.108.107401} {\bibfield  {journal} {\bibinfo  {journal}
  {Physical Review Letters}\ }\textbf {\bibinfo {volume} {108}},\ \bibinfo
  {pages} {107401} (\bibinfo {year} {2012})}\BibitemShut {NoStop}%
\bibitem [{\citenamefont {Berthelot}\ \emph {et~al.}(2006)\citenamefont
  {Berthelot}, \citenamefont {Favero}, \citenamefont {Cassabois}, \citenamefont
  {Voisin}, \citenamefont {Delalande}, \citenamefont {Roussignol},
  \citenamefont {Ferreira},\ and\ \citenamefont
  {G{\'e}rard}}]{berthelot2006unconventional}%
  \BibitemOpen
  \bibfield  {author} {\bibinfo {author} {\bibfnamefont {A.}~\bibnamefont
  {Berthelot}}, \bibinfo {author} {\bibfnamefont {I.}~\bibnamefont {Favero}},
  \bibinfo {author} {\bibfnamefont {G.}~\bibnamefont {Cassabois}}, \bibinfo
  {author} {\bibfnamefont {C.}~\bibnamefont {Voisin}}, \bibinfo {author}
  {\bibfnamefont {C.}~\bibnamefont {Delalande}}, \bibinfo {author}
  {\bibfnamefont {P.}~\bibnamefont {Roussignol}}, \bibinfo {author}
  {\bibfnamefont {R.}~\bibnamefont {Ferreira}}, \ and\ \bibinfo {author}
  {\bibfnamefont {J.-M.}\ \bibnamefont {G{\'e}rard}},\ }\href@noop {}
  {\bibfield  {journal} {\bibinfo  {journal} {Nature Physics}\ }\textbf
  {\bibinfo {volume} {2}},\ \bibinfo {pages} {759} (\bibinfo {year}
  {2006})}\BibitemShut {NoStop}%
\bibitem [{\citenamefont {Bounouar}\ \emph {et~al.}(2012)\citenamefont
  {Bounouar}, \citenamefont {Trichet}, \citenamefont {Elouneg-Jamroz},
  \citenamefont {Andr{\'e}}, \citenamefont {Bellet-Amalric}, \citenamefont
  {Bougerol}, \citenamefont {Den~Hertog}, \citenamefont {Kheng}, \citenamefont
  {Tatarenko},\ and\ \citenamefont {Poizat}}]{bounouar2012extraction}%
  \BibitemOpen
  \bibfield  {author} {\bibinfo {author} {\bibfnamefont {S.}~\bibnamefont
  {Bounouar}}, \bibinfo {author} {\bibfnamefont {A.}~\bibnamefont {Trichet}},
  \bibinfo {author} {\bibfnamefont {M.}~\bibnamefont {Elouneg-Jamroz}},
  \bibinfo {author} {\bibfnamefont {R.}~\bibnamefont {Andr{\'e}}}, \bibinfo
  {author} {\bibfnamefont {E.}~\bibnamefont {Bellet-Amalric}}, \bibinfo
  {author} {\bibfnamefont {C.}~\bibnamefont {Bougerol}}, \bibinfo {author}
  {\bibfnamefont {M.}~\bibnamefont {Den~Hertog}}, \bibinfo {author}
  {\bibfnamefont {K.}~\bibnamefont {Kheng}}, \bibinfo {author} {\bibfnamefont
  {S.}~\bibnamefont {Tatarenko}}, \ and\ \bibinfo {author} {\bibfnamefont
  {J.-P.}\ \bibnamefont {Poizat}},\ }\href@noop {} {\bibfield  {journal}
  {\bibinfo  {journal} {Physical Review B}\ }\textbf {\bibinfo {volume} {86}},\
  \bibinfo {pages} {085325} (\bibinfo {year} {2012})}\BibitemShut {NoStop}%
\bibitem [{\citenamefont {Wang}\ \emph {et~al.}(2004)\citenamefont {Wang},
  \citenamefont {Badolato}, \citenamefont {Wilson-Rae}, \citenamefont
  {Petroff}, \citenamefont {Hu}, \citenamefont {Urayama},\ and\ \citenamefont
  {Imamo{\u{g}}lu}}]{wang2004optical}%
  \BibitemOpen
  \bibfield  {author} {\bibinfo {author} {\bibfnamefont {C.}~\bibnamefont
  {Wang}}, \bibinfo {author} {\bibfnamefont {A.}~\bibnamefont {Badolato}},
  \bibinfo {author} {\bibfnamefont {I.}~\bibnamefont {Wilson-Rae}}, \bibinfo
  {author} {\bibfnamefont {P.}~\bibnamefont {Petroff}}, \bibinfo {author}
  {\bibfnamefont {E.}~\bibnamefont {Hu}}, \bibinfo {author} {\bibfnamefont
  {J.}~\bibnamefont {Urayama}}, \ and\ \bibinfo {author} {\bibfnamefont
  {A.}~\bibnamefont {Imamo{\u{g}}lu}},\ }\href@noop {} {\bibfield  {journal}
  {\bibinfo  {journal} {Applied Physics Letters}\ }\textbf {\bibinfo {volume}
  {85}},\ \bibinfo {pages} {3423} (\bibinfo {year} {2004})}\BibitemShut
  {NoStop}%
\bibitem [{\citenamefont {Davan\ifmmode~\mbox{\c{c}}\else \c{c}\fi{}o}\ \emph
  {et~al.}(2014)\citenamefont {Davan\ifmmode~\mbox{\c{c}}\else \c{c}\fi{}o},
  \citenamefont {Hellberg}, \citenamefont {Ates}, \citenamefont {Badolato},\
  and\ \citenamefont {Srinivasan}}]{PhysRevB.89.161303}%
  \BibitemOpen
  \bibfield  {author} {\bibinfo {author} {\bibfnamefont {M.}~\bibnamefont
  {Davan\ifmmode~\mbox{\c{c}}\else \c{c}\fi{}o}}, \bibinfo {author}
  {\bibfnamefont {C.~S.}\ \bibnamefont {Hellberg}}, \bibinfo {author}
  {\bibfnamefont {S.}~\bibnamefont {Ates}}, \bibinfo {author} {\bibfnamefont
  {A.}~\bibnamefont {Badolato}}, \ and\ \bibinfo {author} {\bibfnamefont
  {K.}~\bibnamefont {Srinivasan}},\ }\href {\doibase
  10.1103/PhysRevB.89.161303} {\bibfield  {journal} {\bibinfo  {journal} {Phys.
  Rev. B}\ }\textbf {\bibinfo {volume} {89}},\ \bibinfo {pages} {161303}
  (\bibinfo {year} {2014})}\BibitemShut {NoStop}%
\bibitem [{\citenamefont {Yeo}\ \emph {et~al.}(2011)\citenamefont {Yeo},
  \citenamefont {Malik}, \citenamefont {Munsch}, \citenamefont {Dupuy},
  \citenamefont {Bleuse}, \citenamefont {Niquet}, \citenamefont {G\'{e}rard},
  \citenamefont {Claudon}, \citenamefont {Wagner}, \citenamefont {Seidelin},
  \citenamefont {Auff\`{e}ves}, \citenamefont {Poizat},\ and\ \citenamefont
  {Nogues}}]{yeo2011surface}%
  \BibitemOpen
  \bibfield  {author} {\bibinfo {author} {\bibfnamefont {I.}~\bibnamefont
  {Yeo}}, \bibinfo {author} {\bibfnamefont {N.~S.}\ \bibnamefont {Malik}},
  \bibinfo {author} {\bibfnamefont {M.}~\bibnamefont {Munsch}}, \bibinfo
  {author} {\bibfnamefont {E.}~\bibnamefont {Dupuy}}, \bibinfo {author}
  {\bibfnamefont {J.}~\bibnamefont {Bleuse}}, \bibinfo {author} {\bibfnamefont
  {Y.~M.}\ \bibnamefont {Niquet}}, \bibinfo {author} {\bibfnamefont {J.~M.}\
  \bibnamefont {G\'{e}rard}}, \bibinfo {author} {\bibfnamefont
  {J.}~\bibnamefont {Claudon}}, \bibinfo {author} {\bibfnamefont
  {E.}~\bibnamefont {Wagner}}, \bibinfo {author} {\bibfnamefont
  {S.}~\bibnamefont {Seidelin}}, \bibinfo {author} {\bibfnamefont
  {A.}~\bibnamefont {Auff\`{e}ves}}, \bibinfo {author} {\bibfnamefont {J.~P.}\
  \bibnamefont {Poizat}}, \ and\ \bibinfo {author} {\bibfnamefont
  {G.}~\bibnamefont {Nogues}},\ }\href {\doibase 10.1063/1.3665629} {\bibfield
  {journal} {\bibinfo  {journal} {Applied Physics Letters}\ }\textbf {\bibinfo
  {volume} {99}},\ \bibinfo {pages} {233106} (\bibinfo {year}
  {2011})}\BibitemShut {NoStop}%
\bibitem [{\citenamefont {Reimer}\ \emph {et~al.}(2014)\citenamefont {Reimer},
  \citenamefont {Bulgarini}, \citenamefont {Heeres}, \citenamefont {Witek},
  \citenamefont {Versteegh}, \citenamefont {Dalacu}, \citenamefont {Lapointe},
  \citenamefont {Poole},\ and\ \citenamefont {Zwiller}}]{reimer2014overcoming}%
  \BibitemOpen
  \bibfield  {author} {\bibinfo {author} {\bibfnamefont {M.~E.}\ \bibnamefont
  {Reimer}}, \bibinfo {author} {\bibfnamefont {G.}~\bibnamefont {Bulgarini}},
  \bibinfo {author} {\bibfnamefont {R.~W.}\ \bibnamefont {Heeres}}, \bibinfo
  {author} {\bibfnamefont {B.~J.}\ \bibnamefont {Witek}}, \bibinfo {author}
  {\bibfnamefont {M.~A.}\ \bibnamefont {Versteegh}}, \bibinfo {author}
  {\bibfnamefont {D.}~\bibnamefont {Dalacu}}, \bibinfo {author} {\bibfnamefont
  {J.}~\bibnamefont {Lapointe}}, \bibinfo {author} {\bibfnamefont {P.~J.}\
  \bibnamefont {Poole}}, \ and\ \bibinfo {author} {\bibfnamefont
  {V.}~\bibnamefont {Zwiller}},\ }\href@noop {} {\bibfield  {journal} {\bibinfo
   {journal} {arXiv preprint arXiv:1407.2833}\ } (\bibinfo {year}
  {2014})}\BibitemShut {NoStop}%
\bibitem [{\citenamefont {Huber}\ \emph {et~al.}(2014)\citenamefont {Huber},
  \citenamefont {Predojevi{\'c}}, \citenamefont {Khoshnegar}, \citenamefont
  {Dalacu}, \citenamefont {Poole}, \citenamefont {Majedi},\ and\ \citenamefont
  {Weihs}}]{huber2014polarization}%
  \BibitemOpen
  \bibfield  {author} {\bibinfo {author} {\bibfnamefont {T.}~\bibnamefont
  {Huber}}, \bibinfo {author} {\bibfnamefont {A.}~\bibnamefont
  {Predojevi{\'c}}}, \bibinfo {author} {\bibfnamefont {M.}~\bibnamefont
  {Khoshnegar}}, \bibinfo {author} {\bibfnamefont {D.}~\bibnamefont {Dalacu}},
  \bibinfo {author} {\bibfnamefont {P.~J.}\ \bibnamefont {Poole}}, \bibinfo
  {author} {\bibfnamefont {H.}~\bibnamefont {Majedi}}, \ and\ \bibinfo {author}
  {\bibfnamefont {G.}~\bibnamefont {Weihs}},\ }\href@noop {} {\bibfield
  {journal} {\bibinfo  {journal} {arXiv preprint arXiv:1405.3765}\ } (\bibinfo
  {year} {2014})}\BibitemShut {NoStop}%
\bibitem [{\citenamefont {Versteegh}\ \emph {et~al.}(2014)\citenamefont
  {Versteegh}, \citenamefont {Reimer}, \citenamefont {J{\"o}ns}, \citenamefont
  {Dalacu}, \citenamefont {Poole}, \citenamefont {Gulinatti}, \citenamefont
  {Giudice},\ and\ \citenamefont {Zwiller}}]{versteegh2014polarization}%
  \BibitemOpen
  \bibfield  {author} {\bibinfo {author} {\bibfnamefont {M.~A.}\ \bibnamefont
  {Versteegh}}, \bibinfo {author} {\bibfnamefont {M.~E.}\ \bibnamefont
  {Reimer}}, \bibinfo {author} {\bibfnamefont {K.~D.}\ \bibnamefont
  {J{\"o}ns}}, \bibinfo {author} {\bibfnamefont {D.}~\bibnamefont {Dalacu}},
  \bibinfo {author} {\bibfnamefont {P.~J.}\ \bibnamefont {Poole}}, \bibinfo
  {author} {\bibfnamefont {A.}~\bibnamefont {Gulinatti}}, \bibinfo {author}
  {\bibfnamefont {A.}~\bibnamefont {Giudice}}, \ and\ \bibinfo {author}
  {\bibfnamefont {V.}~\bibnamefont {Zwiller}},\ }\href@noop {} {\bibfield
  {journal} {\bibinfo  {journal} {arXiv preprint arXiv:1405.4278}\ } (\bibinfo
  {year} {2014})}\BibitemShut {NoStop}%
\bibitem [{Note1()}]{Note1}%
  \BibitemOpen
  \bibinfo {note} {See Supplemental Material at [URL will be inserted by
  Publisher] for details on the fabrication procedure, determination of the
  experimental photon extraction efficiency as well as FDTD simulation of the
  structure.}\BibitemShut {Stop}%
\bibitem [{\citenamefont {Herklotz}\ \emph {et~al.}(2010)\citenamefont
  {Herklotz}, \citenamefont {Plumhof}, \citenamefont {Rastelli}, \citenamefont
  {Schmidt}, \citenamefont {Schultz},\ and\ \citenamefont
  {Dorr}}]{herklotz2010electrical}%
  \BibitemOpen
  \bibfield  {author} {\bibinfo {author} {\bibfnamefont {A.}~\bibnamefont
  {Herklotz}}, \bibinfo {author} {\bibfnamefont {J.~D.}\ \bibnamefont
  {Plumhof}}, \bibinfo {author} {\bibfnamefont {A.}~\bibnamefont {Rastelli}},
  \bibinfo {author} {\bibfnamefont {O.~G.}\ \bibnamefont {Schmidt}}, \bibinfo
  {author} {\bibfnamefont {L.}~\bibnamefont {Schultz}}, \ and\ \bibinfo
  {author} {\bibfnamefont {K.}~\bibnamefont {Dorr}},\ }\href@noop {} {\bibfield
   {journal} {\bibinfo  {journal} {Journal of Applied Physics}\ }\textbf
  {\bibinfo {volume} {108}},\ \bibinfo {pages} {094101} (\bibinfo {year}
  {2010})}\BibitemShut {NoStop}%
\bibitem [{\citenamefont {Kumar}\ \emph {et~al.}(2011)\citenamefont {Kumar},
  \citenamefont {Trotta}, \citenamefont {Zallo}, \citenamefont {Plumhof},
  \citenamefont {Atkinson}, \citenamefont {Rastelli},\ and\ \citenamefont
  {Schmidt}}]{kumar2011strain}%
  \BibitemOpen
  \bibfield  {author} {\bibinfo {author} {\bibfnamefont {S.}~\bibnamefont
  {Kumar}}, \bibinfo {author} {\bibfnamefont {R.}~\bibnamefont {Trotta}},
  \bibinfo {author} {\bibfnamefont {E.}~\bibnamefont {Zallo}}, \bibinfo
  {author} {\bibfnamefont {J.}~\bibnamefont {Plumhof}}, \bibinfo {author}
  {\bibfnamefont {P.}~\bibnamefont {Atkinson}}, \bibinfo {author}
  {\bibfnamefont {A.}~\bibnamefont {Rastelli}}, \ and\ \bibinfo {author}
  {\bibfnamefont {O.}~\bibnamefont {Schmidt}},\ }\href@noop {} {\bibfield
  {journal} {\bibinfo  {journal} {Applied Physics Letters}\ }\textbf {\bibinfo
  {volume} {99}},\ \bibinfo {pages} {161118} (\bibinfo {year}
  {2011})}\BibitemShut {NoStop}%
\bibitem [{\citenamefont {Dalgarno}\ \emph {et~al.}(2008)\citenamefont
  {Dalgarno}, \citenamefont {Smith}, \citenamefont {McFarlane}, \citenamefont
  {Gerardot}, \citenamefont {Karrai}, \citenamefont {Badolato}, \citenamefont
  {Petroff},\ and\ \citenamefont {Warburton}}]{PhysRevB.77.245311}%
  \BibitemOpen
  \bibfield  {author} {\bibinfo {author} {\bibfnamefont {P.~A.}\ \bibnamefont
  {Dalgarno}}, \bibinfo {author} {\bibfnamefont {J.~M.}\ \bibnamefont {Smith}},
  \bibinfo {author} {\bibfnamefont {J.}~\bibnamefont {McFarlane}}, \bibinfo
  {author} {\bibfnamefont {B.~D.}\ \bibnamefont {Gerardot}}, \bibinfo {author}
  {\bibfnamefont {K.}~\bibnamefont {Karrai}}, \bibinfo {author} {\bibfnamefont
  {A.}~\bibnamefont {Badolato}}, \bibinfo {author} {\bibfnamefont {P.~M.}\
  \bibnamefont {Petroff}}, \ and\ \bibinfo {author} {\bibfnamefont {R.~J.}\
  \bibnamefont {Warburton}},\ }\href {\doibase 10.1103/PhysRevB.77.245311}
  {\bibfield  {journal} {\bibinfo  {journal} {Phys. Rev. B}\ }\textbf {\bibinfo
  {volume} {77}},\ \bibinfo {pages} {245311} (\bibinfo {year}
  {2008})}\BibitemShut {NoStop}%
\bibitem [{\citenamefont {Regelman}\ \emph {et~al.}(2001)\citenamefont
  {Regelman}, \citenamefont {Mizrahi}, \citenamefont {Gershoni}, \citenamefont
  {Ehrenfreund}, \citenamefont {Schoenfeld},\ and\ \citenamefont
  {Petroff}}]{PhysRevLett.87.257401}%
  \BibitemOpen
  \bibfield  {author} {\bibinfo {author} {\bibfnamefont {D.~V.}\ \bibnamefont
  {Regelman}}, \bibinfo {author} {\bibfnamefont {U.}~\bibnamefont {Mizrahi}},
  \bibinfo {author} {\bibfnamefont {D.}~\bibnamefont {Gershoni}}, \bibinfo
  {author} {\bibfnamefont {E.}~\bibnamefont {Ehrenfreund}}, \bibinfo {author}
  {\bibfnamefont {W.~V.}\ \bibnamefont {Schoenfeld}}, \ and\ \bibinfo {author}
  {\bibfnamefont {P.~M.}\ \bibnamefont {Petroff}},\ }\href {\doibase
  10.1103/PhysRevLett.87.257401} {\bibfield  {journal} {\bibinfo  {journal}
  {Physical Review Letters}\ }\textbf {\bibinfo {volume} {87}},\ \bibinfo
  {pages} {257401} (\bibinfo {year} {2001})}\BibitemShut {NoStop}%
\bibitem [{\citenamefont {Gerardot}\ \emph {et~al.}(2005)\citenamefont
  {Gerardot}, \citenamefont {Strauf}, \citenamefont {de~Dood}, \citenamefont
  {Bychkov}, \citenamefont {Badolato}, \citenamefont {Hennessy}, \citenamefont
  {Hu}, \citenamefont {Bouwmeester},\ and\ \citenamefont
  {Petroff}}]{gerardot2005coupledqds}%
  \BibitemOpen
  \bibfield  {author} {\bibinfo {author} {\bibfnamefont {B.~D.}\ \bibnamefont
  {Gerardot}}, \bibinfo {author} {\bibfnamefont {S.}~\bibnamefont {Strauf}},
  \bibinfo {author} {\bibfnamefont {M.~J.~A.}\ \bibnamefont {de~Dood}},
  \bibinfo {author} {\bibfnamefont {A.~M.}\ \bibnamefont {Bychkov}}, \bibinfo
  {author} {\bibfnamefont {A.}~\bibnamefont {Badolato}}, \bibinfo {author}
  {\bibfnamefont {K.}~\bibnamefont {Hennessy}}, \bibinfo {author}
  {\bibfnamefont {E.~L.}\ \bibnamefont {Hu}}, \bibinfo {author} {\bibfnamefont
  {D.}~\bibnamefont {Bouwmeester}}, \ and\ \bibinfo {author} {\bibfnamefont
  {P.~M.}\ \bibnamefont {Petroff}},\ }\href@noop {} {\bibfield  {journal}
  {\bibinfo  {journal} {Physical Review Letters}\ }\textbf {\bibinfo {volume}
  {95}},\ \bibinfo {pages} {137403} (\bibinfo {year} {2005})}\BibitemShut
  {NoStop}%
\bibitem [{\citenamefont {Santori}\ \emph {et~al.}(2004)\citenamefont
  {Santori}, \citenamefont {Fattal}, \citenamefont {Vu\ifmmode \check{c}\else
  \v{c}\fi{}kovi\ifmmode~\acute{c}\else \'{c}\fi{}}, \citenamefont {Solomon},
  \citenamefont {Waks},\ and\ \citenamefont {Yamamoto}}]{PhysRevB.69.205324}%
  \BibitemOpen
  \bibfield  {author} {\bibinfo {author} {\bibfnamefont {C.}~\bibnamefont
  {Santori}}, \bibinfo {author} {\bibfnamefont {D.}~\bibnamefont {Fattal}},
  \bibinfo {author} {\bibfnamefont {J.}~\bibnamefont {Vu\ifmmode \check{c}\else
  \v{c}\fi{}kovi\ifmmode~\acute{c}\else \'{c}\fi{}}}, \bibinfo {author}
  {\bibfnamefont {G.~S.}\ \bibnamefont {Solomon}}, \bibinfo {author}
  {\bibfnamefont {E.}~\bibnamefont {Waks}}, \ and\ \bibinfo {author}
  {\bibfnamefont {Y.}~\bibnamefont {Yamamoto}},\ }\href {\doibase
  10.1103/PhysRevB.69.205324} {\bibfield  {journal} {\bibinfo  {journal} {Phys.
  Rev. B}\ }\textbf {\bibinfo {volume} {69}},\ \bibinfo {pages} {205324}
  (\bibinfo {year} {2004})}\BibitemShut {NoStop}%
\bibitem [{\citenamefont {Vamivakas}\ \emph {et~al.}(2011)\citenamefont
  {Vamivakas}, \citenamefont {Zhao}, \citenamefont {F\"alt}, \citenamefont
  {Badolato}, \citenamefont {Taylor},\ and\ \citenamefont
  {Atat\"ure}}]{PhysRevLett.107.166802}%
  \BibitemOpen
  \bibfield  {author} {\bibinfo {author} {\bibfnamefont {A.~N.}\ \bibnamefont
  {Vamivakas}}, \bibinfo {author} {\bibfnamefont {Y.}~\bibnamefont {Zhao}},
  \bibinfo {author} {\bibfnamefont {S.}~\bibnamefont {F\"alt}}, \bibinfo
  {author} {\bibfnamefont {A.}~\bibnamefont {Badolato}}, \bibinfo {author}
  {\bibfnamefont {J.~M.}\ \bibnamefont {Taylor}}, \ and\ \bibinfo {author}
  {\bibfnamefont {M.}~\bibnamefont {Atat\"ure}},\ }\href {\doibase
  10.1103/PhysRevLett.107.166802} {\bibfield  {journal} {\bibinfo  {journal}
  {Physical Review Letters}\ }\textbf {\bibinfo {volume} {107}},\ \bibinfo
  {pages} {166802} (\bibinfo {year} {2011})}\BibitemShut {NoStop}%
\bibitem [{\citenamefont {J{\"o}ns}\ \emph {et~al.}(2011)\citenamefont
  {J{\"o}ns}, \citenamefont {Hafenbrak}, \citenamefont {Singh}, \citenamefont
  {Ding}, \citenamefont {Plumhof}, \citenamefont {Rastelli}, \citenamefont
  {Schmidt}, \citenamefont {Bester},\ and\ \citenamefont
  {Michler}}]{jons2011dependence}%
  \BibitemOpen
  \bibfield  {author} {\bibinfo {author} {\bibfnamefont {K.}~\bibnamefont
  {J{\"o}ns}}, \bibinfo {author} {\bibfnamefont {R.}~\bibnamefont {Hafenbrak}},
  \bibinfo {author} {\bibfnamefont {R.}~\bibnamefont {Singh}}, \bibinfo
  {author} {\bibfnamefont {F.}~\bibnamefont {Ding}}, \bibinfo {author}
  {\bibfnamefont {J.}~\bibnamefont {Plumhof}}, \bibinfo {author} {\bibfnamefont
  {A.}~\bibnamefont {Rastelli}}, \bibinfo {author} {\bibfnamefont
  {O.}~\bibnamefont {Schmidt}}, \bibinfo {author} {\bibfnamefont
  {G.}~\bibnamefont {Bester}}, \ and\ \bibinfo {author} {\bibfnamefont
  {P.}~\bibnamefont {Michler}},\ }\href@noop {} {\bibfield  {journal} {\bibinfo
   {journal} {Physical Review Letters}\ }\textbf {\bibinfo {volume} {107}},\
  \bibinfo {pages} {217402} (\bibinfo {year} {2011})}\BibitemShut {NoStop}%
\end{thebibliography}

\begin{thebibliography}{4}%
\makeatletter
\providecommand \@ifxundefined [1]{%
 \@ifx{#1\undefined}
}%
\providecommand \@ifnum [1]{%
 \ifnum #1\expandafter \@firstoftwo
 \else \expandafter \@secondoftwo
 \fi
}%
\providecommand \@ifx [1]{%
 \ifx #1\expandafter \@firstoftwo
 \else \expandafter \@secondoftwo
 \fi
}%
\providecommand \natexlab [1]{#1}%
\providecommand \enquote  [1]{``#1''}%
\providecommand \bibnamefont  [1]{#1}%
\providecommand \bibfnamefont [1]{#1}%
\providecommand \citenamefont [1]{#1}%
\providecommand \href@noop [0]{\@secondoftwo}%
\providecommand \href [0]{\begingroup \@sanitize@url \@href}%
\providecommand \@href[1]{\@@startlink{#1}\@@href}%
\providecommand \@@href[1]{\endgroup#1\@@endlink}%
\providecommand \@sanitize@url [0]{\catcode `\\12\catcode `\$12\catcode
  `\&12\catcode `\#12\catcode `\^12\catcode `\_12\catcode `\%12\relax}%
\providecommand \@@startlink[1]{}%
\providecommand \@@endlink[0]{}%
\providecommand \url  [0]{\begingroup\@sanitize@url \@url }%
\providecommand \@url [1]{\endgroup\@href {#1}{\urlprefix }}%
\providecommand \urlprefix  [0]{URL }%
\providecommand \Eprint [0]{\href }%
\providecommand \doibase [0]{http://dx.doi.org/}%
\providecommand \selectlanguage [0]{\@gobble}%
\providecommand \bibinfo  [0]{\@secondoftwo}%
\providecommand \bibfield  [0]{\@secondoftwo}%
\providecommand \translation [1]{[#1]}%
\providecommand \BibitemOpen [0]{}%
\providecommand \bibitemStop [0]{}%
\providecommand \bibitemNoStop [0]{.\EOS\space}%
\providecommand \EOS [0]{\spacefactor3000\relax}%
\providecommand \BibitemShut  [1]{\csname bibitem#1\endcsname}%
\let\auto@bib@innerbib\@empty
\bibitem [{\citenamefont {Friedler}\ \emph {et~al.}(2009)\citenamefont
  {Friedler}, \citenamefont {Sauvan}, \citenamefont {Hugonin}, \citenamefont
  {Lalanne}, \citenamefont {Claudon},\ and\ \citenamefont
  {G{\'e}rard}}]{friedler2009solid}%
  \BibitemOpen
  \bibfield  {author} {\bibinfo {author} {\bibfnamefont {I.}~\bibnamefont
  {Friedler}}, \bibinfo {author} {\bibfnamefont {C.}~\bibnamefont {Sauvan}},
  \bibinfo {author} {\bibfnamefont {J.-P.}\ \bibnamefont {Hugonin}}, \bibinfo
  {author} {\bibfnamefont {P.}~\bibnamefont {Lalanne}}, \bibinfo {author}
  {\bibfnamefont {J.}~\bibnamefont {Claudon}}, \ and\ \bibinfo {author}
  {\bibfnamefont {J.-M.}\ \bibnamefont {G{\'e}rard}},\ }\href@noop {}
  {\bibfield  {journal} {\bibinfo  {journal} {Optics Express}\ }\textbf
  {\bibinfo {volume} {17}},\ \bibinfo {pages} {2095} (\bibinfo {year}
  {2009})}\BibitemShut {NoStop}%
\bibitem [{\citenamefont {Santori}\ \emph {et~al.}(2004)\citenamefont
  {Santori}, \citenamefont {Fattal}, \citenamefont {Vu\ifmmode \check{c}\else
  \v{c}\fi{}kovi\ifmmode~\acute{c}\else \'{c}\fi{}}, \citenamefont {Solomon},
  \citenamefont {Waks},\ and\ \citenamefont {Yamamoto}}]{PhysRevB.69.205324}%
  \BibitemOpen
  \bibfield  {author} {\bibinfo {author} {\bibfnamefont {C.}~\bibnamefont
  {Santori}}, \bibinfo {author} {\bibfnamefont {D.}~\bibnamefont {Fattal}},
  \bibinfo {author} {\bibfnamefont {J.}~\bibnamefont {Vu\ifmmode \check{c}\else
  \v{c}\fi{}kovi\ifmmode~\acute{c}\else \'{c}\fi{}}}, \bibinfo {author}
  {\bibfnamefont {G.~S.}\ \bibnamefont {Solomon}}, \bibinfo {author}
  {\bibfnamefont {E.}~\bibnamefont {Waks}}, \ and\ \bibinfo {author}
  {\bibfnamefont {Y.}~\bibnamefont {Yamamoto}},\ }\href {\doibase
  10.1103/PhysRevB.69.205324} {\bibfield  {journal} {\bibinfo  {journal} {Phys.
  Rev. B}\ }\textbf {\bibinfo {volume} {69}},\ \bibinfo {pages} {205324}
  (\bibinfo {year} {2004})}\BibitemShut {NoStop}%
\bibitem [{\citenamefont {Davan\ifmmode~\mbox{\c{c}}\else \c{c}\fi{}o}\ \emph
  {et~al.}(2014)\citenamefont {Davan\ifmmode~\mbox{\c{c}}\else \c{c}\fi{}o},
  \citenamefont {Hellberg}, \citenamefont {Ates}, \citenamefont {Badolato},\
  and\ \citenamefont {Srinivasan}}]{PhysRevB.89.161303}%
  \BibitemOpen
  \bibfield  {author} {\bibinfo {author} {\bibfnamefont {M.}~\bibnamefont
  {Davan\ifmmode~\mbox{\c{c}}\else \c{c}\fi{}o}}, \bibinfo {author}
  {\bibfnamefont {C.~S.}\ \bibnamefont {Hellberg}}, \bibinfo {author}
  {\bibfnamefont {S.}~\bibnamefont {Ates}}, \bibinfo {author} {\bibfnamefont
  {A.}~\bibnamefont {Badolato}}, \ and\ \bibinfo {author} {\bibfnamefont
  {K.}~\bibnamefont {Srinivasan}},\ }\href {\doibase
  10.1103/PhysRevB.89.161303} {\bibfield  {journal} {\bibinfo  {journal} {Phys.
  Rev. B}\ }\textbf {\bibinfo {volume} {89}},\ \bibinfo {pages} {161303}
  (\bibinfo {year} {2014})}\BibitemShut {NoStop}%
\bibitem [{\citenamefont {Dalgarno}\ \emph {et~al.}(2008)\citenamefont
  {Dalgarno}, \citenamefont {Smith}, \citenamefont {McFarlane}, \citenamefont
  {Gerardot}, \citenamefont {Karrai}, \citenamefont {Badolato}, \citenamefont
  {Petroff},\ and\ \citenamefont {Warburton}}]{PhysRevB.77.245311}%
  \BibitemOpen
  \bibfield  {author} {\bibinfo {author} {\bibfnamefont {P.~A.}\ \bibnamefont
  {Dalgarno}}, \bibinfo {author} {\bibfnamefont {J.~M.}\ \bibnamefont {Smith}},
  \bibinfo {author} {\bibfnamefont {J.}~\bibnamefont {McFarlane}}, \bibinfo
  {author} {\bibfnamefont {B.~D.}\ \bibnamefont {Gerardot}}, \bibinfo {author}
  {\bibfnamefont {K.}~\bibnamefont {Karrai}}, \bibinfo {author} {\bibfnamefont
  {A.}~\bibnamefont {Badolato}}, \bibinfo {author} {\bibfnamefont {P.~M.}\
  \bibnamefont {Petroff}}, \ and\ \bibinfo {author} {\bibfnamefont {R.~J.}\
  \bibnamefont {Warburton}},\ }\href {\doibase 10.1103/PhysRevB.77.245311}
  {\bibfield  {journal} {\bibinfo  {journal} {Phys. Rev. B}\ }\textbf {\bibinfo
  {volume} {77}},\ \bibinfo {pages} {245311} (\bibinfo {year}
  {2008})}\BibitemShut {NoStop}%
\end{thebibliography}
%


 }

\end{document}